\documentclass[reprint,aps,superscriptaddress]{revtex4-2}
\pdfoutput=1

\usepackage{graphicx}
\usepackage{geometry}
\usepackage{multirow}
\usepackage{lipsum} 
\usepackage{tcolorbox}
\usepackage{bm}
\usepackage{xr}
\usepackage{float}
\usepackage{amsmath}
\usepackage{amssymb}
\usepackage{amsthm}
\usepackage{amsfonts}
\usepackage{booktabs,siunitx}
\usepackage{xcolor}
\usepackage{soul}
\usepackage{cleveref}
\usepackage{svg}
\usepackage[normalem]{ulem}
\usepackage{booktabs}
\usepackage{xspace}

\usepackage{algorithm}
\usepackage{algpseudocode}

\renewcommand{\sout}[1]{}
\algnewcommand\algorithmicforeach{\textbf{for each}}
\algdef{SE}[FOR]{ForEach}{EndForEach}[1]{\algorithmicforeach\ #1\ \algorithmicdo}{\algorithmicend\ \algorithmicforeach}

\makeatletter
\newenvironment{breakablealgorithm}
  {
     \refstepcounter{algorithm}
     \hrule height.8pt depth0pt \kern2pt
     \renewcommand{\caption}[2][\relax]{
       {\raggedright\textbf{\ALG@name~\thealgorithm} ##2\par}%
       \ifx\relax##1\relax 
         \addcontentsline{loa}{algorithm}{\protect\numberline{\thealgorithm}##2}%
       \else 
         \addcontentsline{loa}{algorithm}{\protect\numberline{\thealgorithm}##1}%
       \fi
       \kern2pt\hrule\kern2pt
     }
  }{
     \kern2pt\hrule\relax
  
  }
\makeatother

\newcommand{\bitcoin}{\textsc{Bitcoin}\xspace}

\newcommand{\Citation}{\textsc{Citation}\xspace}

\newcommand{\qna}{\textsc{qna}\xspace}
\newcommand{\Email}{\textsc{E-mail}\xspace}

\newcommand{\metabolic}{\textsc{Metabolic}\xspace}

\usepackage{xcolor}

\begin{document}

\title{The microscale organization of directed hypergraphs}

\author{Quintino Francesco Lotito}
\email{lotitoq@ceu.edu}
\affiliation{Department of Network and Data Science, Central European University, 1100 Vienna, Austria}
\affiliation{Department of Information Engineering and Computer Science, University of Trento, via Sommarive 9, 38123 Trento, Italy}

\author{Alberto Vendramini}
\affiliation{Department of Information Engineering and Computer Science, University of Trento, via Sommarive 9, 38123 Trento, Italy}

\author{Alberto Montresor}
\affiliation{Department of Information Engineering and Computer Science, University of Trento, via Sommarive 9, 38123 Trento, Italy}

\author{Federico Battiston}
\email{battistonf@ceu.edu}
\affiliation{Department of Network and Data Science, Central European University, 1100 Vienna, Austria}

\begin{abstract}
Many real-world complex systems are characterized by non-pairwise -- higher-order -- interactions among system's units, and can be effectively modeled as hypergraphs. Directed hypergraphs distinguish between source and target sets within each hyperedge, and allow to account for the directional flow of information between nodes. Here, we provide a framework to characterize the structural organization of directed higher-order networks at their microscale. First, we extract the fingerprint of a directed hypergraph, capturing the frequency of hyperedges with a certain source and target sizes, and use this information to compute differences in higher-order connectivity patterns among real-world systems. Then, we investigate the overlap among sources and targets to reveal recurring sets of co-sending and co-receiving nodes. We define reciprocity in hypergraphs using exact, strong, and weak definitions to quantify the extent to which hyperedges are reciprocated. Finally, we extend motif analysis to identify recurring interaction patterns and extract the building blocks of directed hypergraphs. We validate our framework on empirical datasets, including Bitcoin transactions, metabolic networks, and citation data, revealing structural principles behind the organization of real-world systems.  
\end{abstract}

\maketitle

\section*{Introduction}
Accurately modeling interactions among entities is crucial to understand the properties of many complex systems. Traditional network models focus on pairwise connections between nodes~\cite{boccaletti2006complex, cimini2019statistical}, neglecting the complexities of systems where multiple units interact simultaneously. Such higher-order interactions are prevalent in various domains, including social networks~\cite{patania2017shape, cencetti2021temporal, iacopini2024temporal}, folksonomies~\cite{ghoshal2009random},  ecological systems~\cite{grilli2017higher}, chemical reactions~\cite{jost2019hypergraph} including metabolic pathways~\cite{traversa2023robustness}, and the brain~\cite{petri2014homological, santoro2023higher}.

Hypergraphs~\cite{berge1973graphs} provide a framework for explicitly encoding higher-order interactions, representing them as hyperedges connecting multiple nodes simultaneously. By preserving group-based interactions, they improve our ability to understand the structures and dynamics of systems with many-body interactions~\cite{battiston2020networks, battiston2021physics}. Recently, a variety of measures have been introduced or extended to capture the higher-order organization of complex systems, including centrality~\cite{benson2019three, tudisco2021node}, community structure~\cite{eriksson2021choosing, contisciani2022inference, ruggeri2023community} and motifs~\cite{lotito2022higher, lee2020hypergraph, arregui2024patterns}. Moreover, new models have allowed to describe systems' evolution~\cite{petri2018simplicial,digaetano2024percolation,gallo2024higher}, and highlight the importance of higher-order interactions in shaping emergent behaviors in
diffusion~\cite{schaub2020random,carletti2020random}, synchronization~\cite{lucas2020multiorder,gambuzza2021stability, zhang2023higher}, spreading~\cite{iacopini2019simplicial, chowdhary2021simplicial} and evolutionary dynamics~\cite{civilini2024explosive}.

Most research has so far focused on undirected hypergraphs, which fail to capture the directional nature of many real-world interactions. For example, in a metabolic reaction, a set of reactants transforms into a set of products~\cite{traversa2023robustness}. Similarly, in a Bitcoin transaction, multiple source wallets may transfer funds simultaneously to multiple target wallets~\cite{ranshous2017exchange}. To accurately encode such interactions, models must incorporate directionality into their representations. In this sense, directed hypergraphs enhance modeling by distinguishing between source and target sets in each hyperedge~\cite{gallo1993directed}. Tools to study directed hypergraphs are largely underdeveloped, with notable exceptions in areas such as null models~\cite{preti2024higher}, synchronization~\cite{gallo2022synchronization}, overlapping patterns between two hyperedges of limited size~\cite{moon2023four}, and some early proposals to define reciprocity~\cite{pearcy2014hypergraph, kim2023reciprocity}.

In this work, we introduce measures and tools to characterize the microscale organization of real-world directed hypergraphs. First, we discuss a decomposition into fundamental interaction types: one-to-one, one-to-many, many-to-one and many-to-many. We analyze empirical data to count the occurrences of each interaction type, and use this information as a signature to compute differences in higher-order connectivity patterns. Then, for each node, we investigate the overlap among its source and target sets, to extract recurring groups of co-senders and co-receivers. By examining this overlap and comparing it against randomised models, we aim to reveal whether certain systems exhibit a more redundant organization, where interactions frequently recur among the same groups, or a more diverse structure with less overlap among participants. Additionally, we propose new, computationally efficient definitions for reciprocity~\cite{garlaschelli2004patterns} for directed hypergraphs, namely exact, strong and weak higher-order reciprocity, designed to capture different patterns of bi-directionality in empirical data. Finally, we extend motif analysis~\cite{milo2002network} to incorporate the directionality of interactions, extracting recurring higher-order and directed subgraphs. 
Our results suggest the existence of complex mechanisms of feedback and reinforcement in the information flow among system units, where pairwise interactions support the action of groups, and vice versa.


\section*{Results}

\begin{figure}
    \centering
    \includegraphics[width=\linewidth]{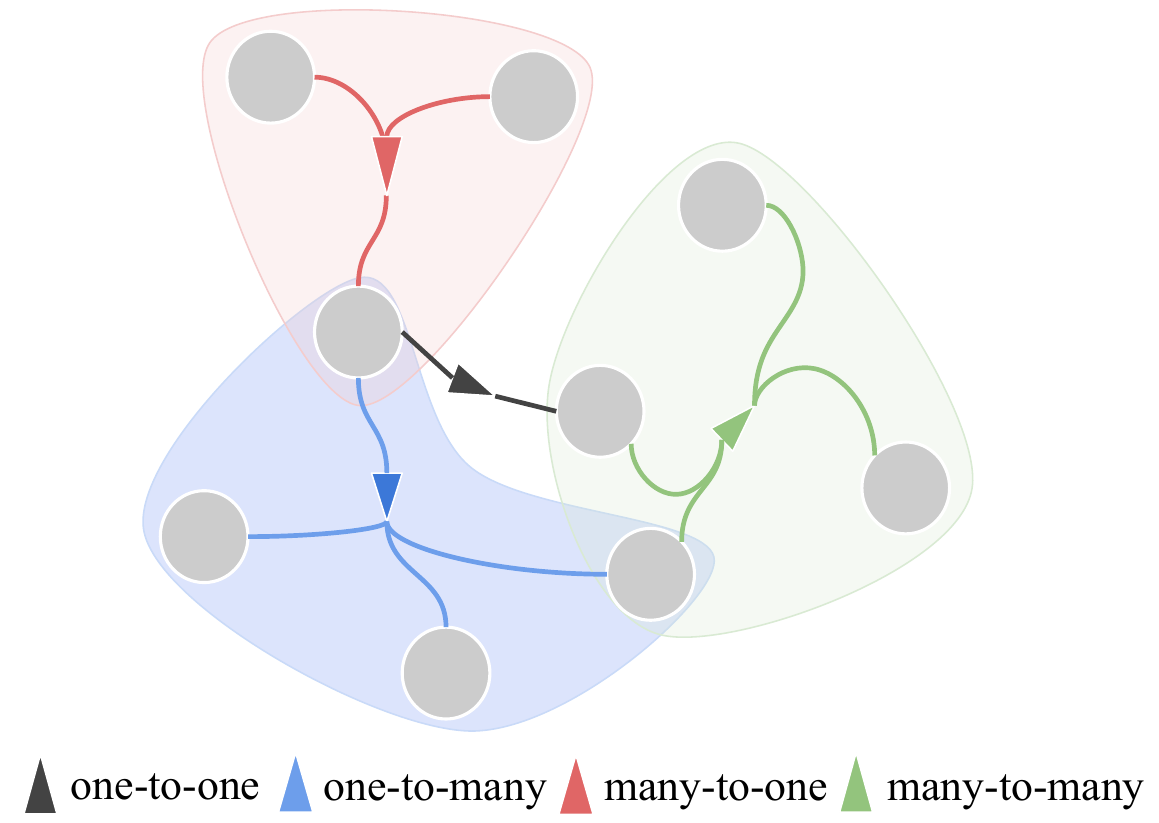}
    \caption{\textbf{Schematic of a directed hypergraph.} Each interaction encodes a source set of units acting towards a target set of units. We distinguish four types of directed higher-order interactions: one-to-one (black), one-to-many (blue), many-to-one (red), and many-to-many (green).}
    \label{fig:fig1}
\end{figure}

Traditional graph models reduce directed group interactions into a collection of pairwise links, often leading to a loss of important structural information about group organization and dynamics. For instance, reducing a many-to-many interaction such as \textsc{SOURCE =} $\{A,B\}$ and \textsc{TARGET =} $\{D,E\}$ to a set of directed pairwise links ($A \to D$, $A \to E$, $B \to D$ and $B \to D$) fails to capture the collective nature of the interaction, including information about co-senders and co-receivers nodes. Directed hypergraphs preserve both group-based structure and the associated information flow, allowing for a more faithful representation of complex interactions. In such a framework, hyperedge direction is encoded by distinguishing between source and target node sets, which we consider non-empty and disjoint. More formally, we work with finite, simple, directed hypergraphs $\mathcal{H}=(V,E,\mathsf{s},\mathsf{t})$ on node set $V$, where each hyperedge $e\in E$ is an ordered pair $(\mathsf{s}(e),\mathsf{t}(e))$ with
$\mathsf{s}(e),\mathsf{t}(e)\subseteq V,\ \mathsf{s}(e)\neq\varnothing,\ \mathsf{t}(e)\neq\varnothing \ 
\text{and }\mathsf{s}(e)\cap\mathsf{t}(e)=\varnothing$.  
This definition yields four canonical directed hyperedge patterns: one-to-one, where a single source node connects to a single target; one-to-many, where one source affects multiple targets; many-to-one, where multiple sources act on a single target; and many-to-many, the most general case, where multiple sources act on multiple targets. The analysis of the interplay and overlap among these building blocks in real-world hypergraphs enables a characterization of their microscale organization. Figure~\ref{fig:fig1} illustrates this taxonomy on a toy directed hypergraph.

We analyzed datasets from multiple domains, including \qna (nodes are users and forum posts are hyperedges), \Email (nodes are users and emails are hyperedges), \bitcoin (nodes are accounts and financial transactions are hyperedges), \metabolic (nodes are genes and metabolic reactions are hyperedges) and \Citation (nodes are authors and hyperedges are paper citations)~\cite{kim2023reciprocity}. Each dataset is encoded as a set-indexed adjacency tensor. In particular, we index the distinct source- and target-sets observed in the data by $\{S_\alpha\}_{\alpha=1}^{p}$ and $\{T_\beta\}_{\beta=1}^{q}$, respectively, and define the set-indexed adjacency tensor $\mathcal{A}\in\{0,1\}^{p\times q}$ by $\mathcal{A}_{\alpha\beta}=1$ if and only if there exists $e\in E$ with $\mathsf{s}(e)=S_\alpha$ and $\mathsf{t}(e)=T_\beta$ (and $\mathcal{A}_{\alpha\beta}=0$ otherwise). Whenever $\mathcal{A}_{\alpha\beta}=1$, we enforce $S_\alpha\neq\varnothing$, $T_\beta\neq\varnothing$ and $S_\alpha\cap T_\beta=\varnothing$.

Detailed descriptions and summary statistics of each dataset are reported in Supplementary Note 1.

\subsection*{Patterns of directed hyperedges}

\begin{figure*}
    \centering
    \includegraphics[width=\linewidth]{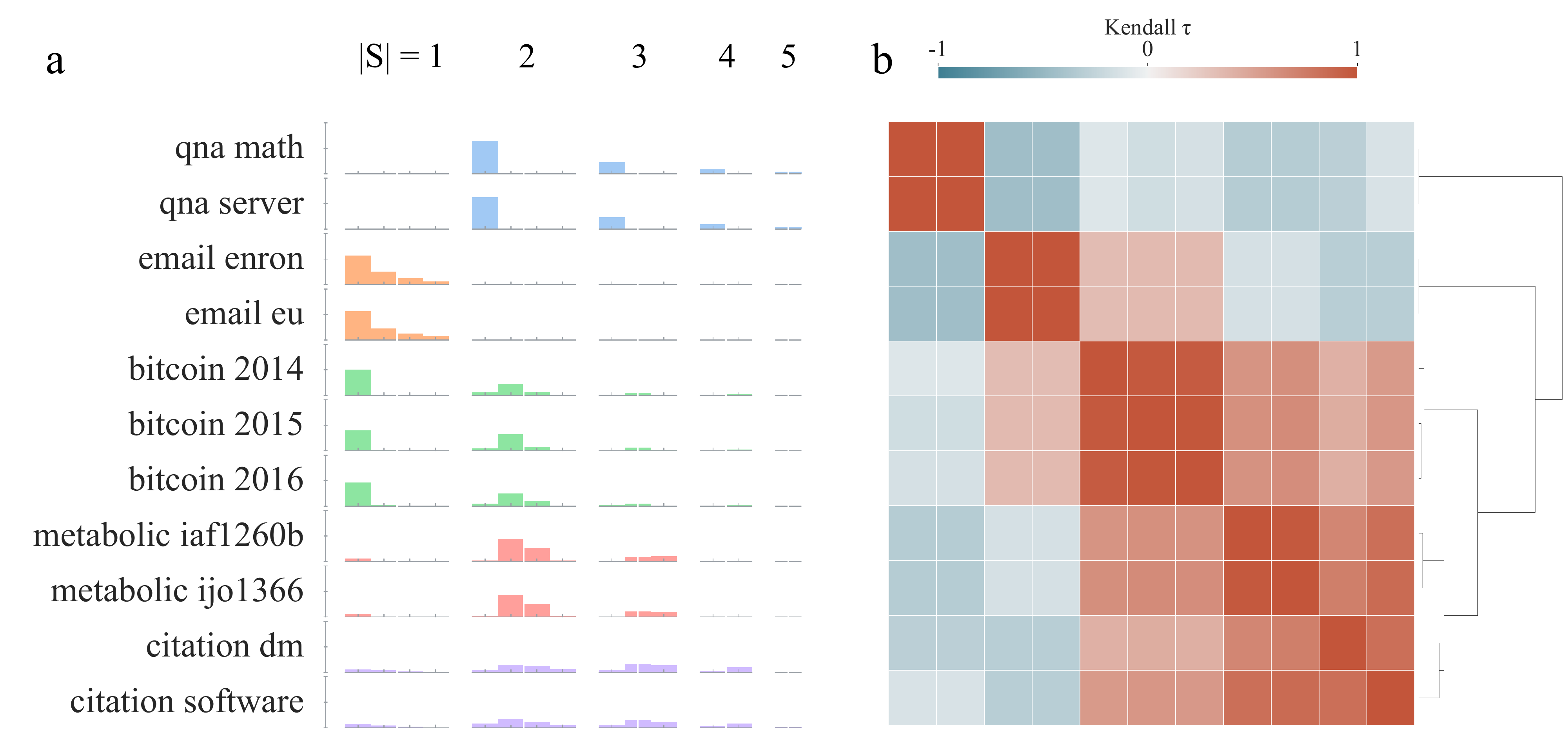}
    \caption{\textbf{Hyperedge signature of directed hypergraphs.} a) We describe each system with a hyperedge signature vector whose entries encode the count of directed hyperedges with source and tail sizes $(|S|,|T|)$. We compute statistics using hyperedges with total cardinality at most $6$ (i.e., $|S|+|T|\le 6$). For visualization, we display each vector as a sequence of histogram panels: one panel for each source size $|S|=i$, separated by a small gap; within each panel, bins correspond to tail sizes $|T|$ in increasing order, restricted by $i+|T|\le 6$. Systems from the same domain share the color. b) Dendrogram resulting from agglomerative clustering applied to the correlation matrix of hyperedge signature vectors for each dataset. Correlation values are color-coded, with high positive correlations in red and high negative correlations in blue.}
    \label{fig:signature}
\end{figure*}

A natural starting point to characterize directed hypergraphs across domains is investigating the diversity in their patterns of directed hyperedges. For each dataset, we construct a \textit{hyperedge signature vector} \( \mathbf{v} \), which captures the distribution of hyperedges based on the sizes of their source and target sets (see Methods). Such vectors provide a fingerprint for systems based on their higher-order connectivity patterns at the microscale. Figure~\ref{fig:signature}a shows the hyperedge signature vectors for each dataset, considering interactions up to size \(6\). To emphasize the role of higher-order interactions in the analysis, we do not consider one-to-one interactions. We find that one-to-many interactions dominate the \Email dataset, reflecting the typical structure of email communications. Similarly, in the \qna, many-to-one interactions are prevalent, as these systems involve multiple individuals responding to a question by a single user. In contrast, \metabolic and \Citation datasets show high abundances in many-to-many relationships across a variety of source and target set sizes. Finally, \bitcoin dataset exhibits more varied behavior, with abundant entries for both one-to-many and many-to-many interactions, indicating different interaction types in the network.

To further explore structural diversity across different domains, we compute pairwise rank correlations among hyperedge signature vectors using weighted Kendall's~$\tau$
and apply hierarchical agglomerative clustering on their correlation matrix. A correlation value close to $1$ indicates similar hyperedge structures, $0$ suggests no relationship, and $-1$ indicates the structures are inversely related. The clustering procedure applied to the systems' correlation matrix results in a dendrogram that visually represents their hierarchical relationships, highlighting the presence of clusters of directed hypergraphs that share similar connectivity patterns. In Figure~\ref{fig:signature}b, we show the correlation matrix and the clustering dendrogram. By examining the correlation matrix, we observe a strong correlation within systems from the same domain, indicating highly similar abundance in hyperedge structures. In contrast, systems from different domains exhibit varying degrees of correlation. Specifically, \Email and \qna datasets are inversely correlated, as they display non-overlapping and complementary connectivity patterns: \Email is characterized by one-to-many interactions, whereas \qna\ primarily involves many-to-one relationships. The \metabolic and \Citation datasets, which feature many-to-many interactions, are positively correlated and form a distinct cluster. Interestingly, the \bitcoin datasets also display positive correlations with the \metabolic and \Citation cluster due to a high presence of many-to-many interaction patterns. However, they also exhibit a weaker positive correlation with the \Email datasets, reflecting the presence of one-to-many interactions in \bitcoin.

\subsection*{Source and target sets overlap}

\begin{figure*}[t]
    \centering
    \includegraphics[width=\linewidth]{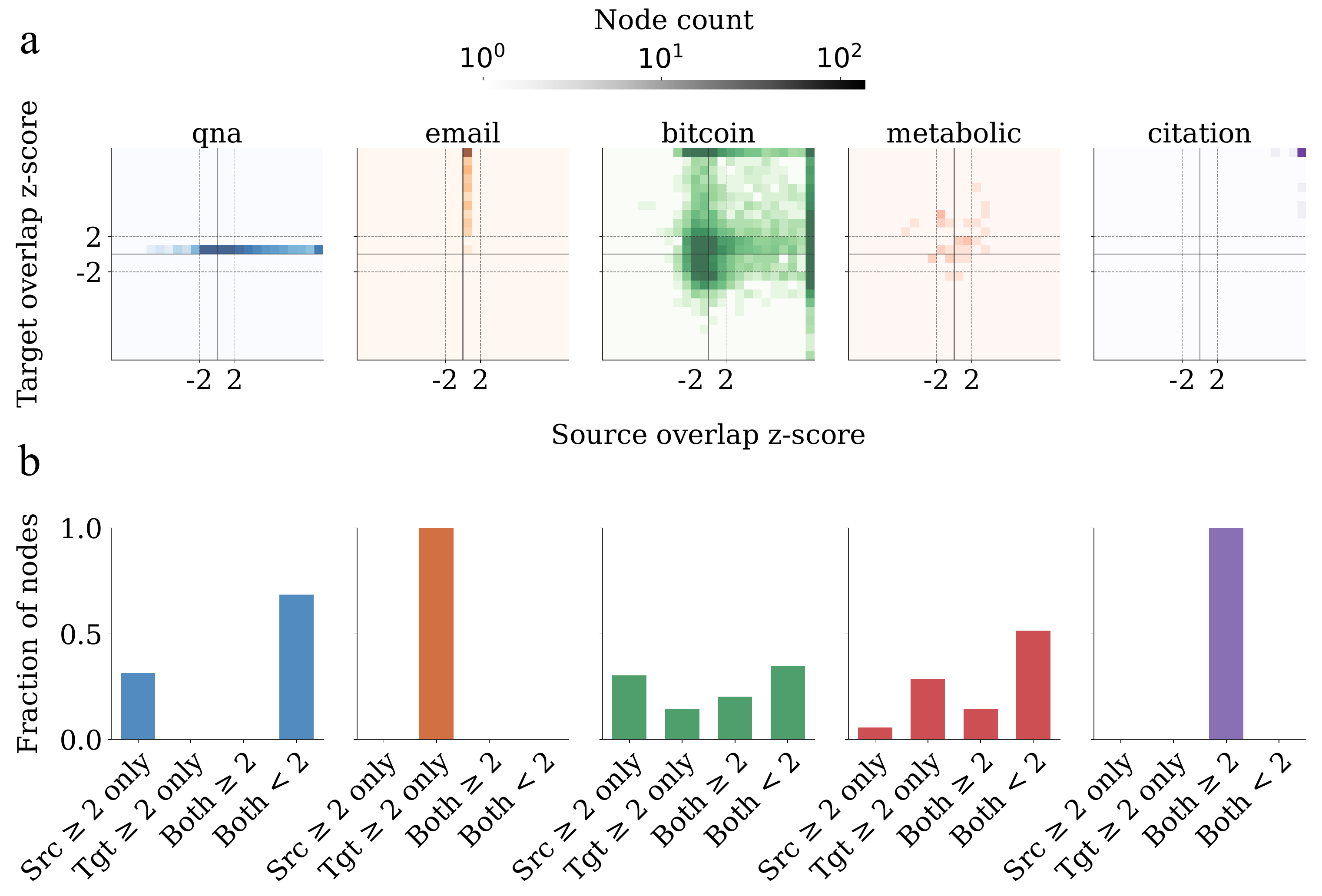}
\caption{\textbf{Overlap across domains.} a) Distribution of node counts within the joint z-score space of source and target overlap, representing how much nodes deviate from null model expectations in both dimensions. b) Bar plots quantifying the fraction of nodes exceeding the $\textit{z-score} \geq 2$ threshold for either source or target overlaps, nodes exceeding both thresholds, and nodes below both thresholds.}
    \label{fig:hyperedge-overlap}
\end{figure*}

The degree to which hyperedges share elements provides valuable insights into redundancy, hierarchical structures, and information flow within the system. In general, real-world systems exhibit a high degree of overlap, indicating that recurrent and redundant interactions are a shared feature~\cite{lee2021hyperedges}. Moreover, hyperedge overlap has been shown to widely impact the dynamics of systems with higher-order interactions~\cite{malizia2025hyperedge, lamata2025hyperedge}.

In directed hypergraphs, nodes are frequently involved in multiple hyperedges, either as part of the source or the target set. In order to characterize and quantify this property, for each node, we measure how much its incident source sets overlap and how much its incident target sets overlap, and we compare with a null model, reporting source and target z-scores (see Methods). In other words, we measure the extent to which nodes engage in interactions with the same set of co-senders or co-receivers. Specifically, an observed overlap associated with a z-score greater than $2$ (i.e., statistically significant excess overlap) highlights nodes that tend to participate in structurally redundant interactions where the same groups of nodes frequently co-occur in source or target sets. Conversely, a z-score less than $-2$ reflects statistically significant lower overlap than expected, suggesting that interactions are more diverse, with hyperedges being more distinct and less likely to share members.

In Fig.~\ref{fig:hyperedge-overlap}, we show the distribution of nodes with a given overlap z-score for source and target sets across domains, along with the fraction of nodes in each region of the z-score space. The focus is on positive excess overlap, as statistically significant negative excess overlap is very rare. The \Citation dataset displays significant overlap for both source and target sets, showing that (i) an author tends to preferentially work with known collaborators, and (ii) an author tends to be cited repeatedly alongside similar sets of authors. In the \Email dataset, the excess overlap can be computed only for the target sets, as the source sets always have cardinality $1$. The overlap is significantly larger than random, underscoring the hierarchical and broadcast-like nature of email communication. The \bitcoin dataset generally exhibits nodes with high excess overlap in both source and target sets. However, the presence of nodes with excess overlap values lower than zero implies that certain participants in the network engage in interactions that introduce more novelty rather than reinforce existing hyperedges. The \metabolic dataset follows a similar trend, with half of the nodes displaying significant excess overlap for either source or target sets or both, suggesting that metabolic reactions tend to involve recurring sets of substrates and products and highlighting the modular nature of metabolic networks.   \qna data show a lower excess overlap compared with the other systems, indicating that forum respondents are less likely to engage repeatedly with the same set of co-responders. Since the target sets in this dataset always have cardinality $1$, the excess overlap can be computed only for source sets.

\subsection*{Higher-order reciprocity}

\begin{figure*}
    \centering
\includegraphics[width=\linewidth]{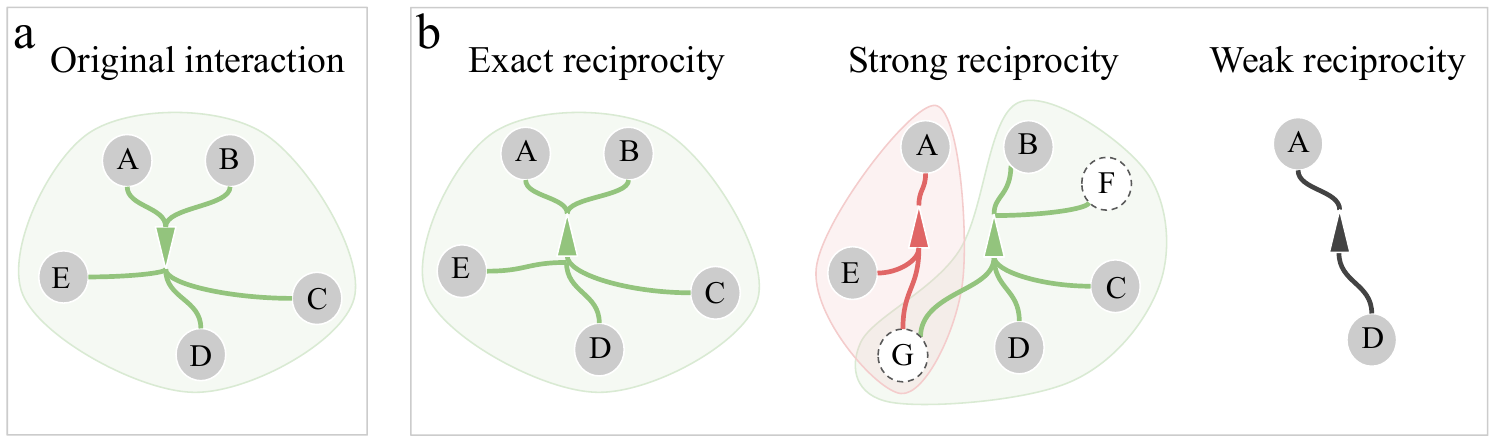}
\caption{\textbf{Reciprocity measures for directed hypergraphs.}
a) Example of a directed hyperedge.
b) All possible ways in which this hyperedge can be reciprocated according to our definitions. Exact reciprocity: a single hyperedge with source and target sets swapped, represented by reversing the arrow between the same node sets. Strong reciprocity: multiple hyperedges collectively reverse the interaction, possibly involving external nodes. Weak reciprocity: at least one node in the target set reciprocates with one node in the source set, illustrated as a pairwise link with reversed direction. In all panels, shaded areas group nodes involved in each interaction; colors encode the interaction pattern (green many-to-many, red many-to-one, black one-to-one); arrows encode direction from source to target. Grey disks denote nodes in the original hyperedge, and external nodes that appear only in reciprocal interactions are white with a dashed border.}
\label{fig:reciprocity_def}
\end{figure*}

Reciprocity is a fundamental property of systems with directed interactions, including social networks~\cite{wasserman1994social}. It traditionally refers to the tendency of the system’s units to mutually exchange information. 
In directed graphs, reciprocity is defined as \[r = \frac{\overleftrightarrow{L}}{L}\], measuring the ratio of the number of bidirectional links ($\overleftrightarrow{L}$) to the total number of links ($L$). This measure is often normalized as \[\rho_{\mathrm{NM}} = \frac{r - \langle r \rangle_{\mathrm{NM}}}{1 - \langle r \rangle_{\mathrm{NM}}}\] where $r$ is the observed reciprocity in the network, and $\langle r \rangle_{\mathrm{NM}}$ is the average reciprocity in null model samples~\cite{garlaschelli2004patterns}. It measures the difference between observed and expected reciprocity by the maximum possible deviation, bounding $\rho_{\mathrm{NM}}$ between $-1$ and $1$. Positive values indicate more reciprocity than expected at random, negative values indicate less, and values near zero suggest consistency with the null model. This normalization allows a more faithful comparison and ranking of reciprocity across systems with different scales and density~\cite{garlaschelli2004patterns}. Recognizing its broad importance, recent works have extended reciprocity to hypergraphs, accounting for the complexity of having multiple nodes in both the source and target sets of hyperedges. Among the recent approaches for hypergraph reciprocity, one method decomposes hyperedges into pairwise links~\cite{pearcy2014hypergraph}, losing information about group interactions. An alternative approach defines a more complex measure that diverges from the traditional binary definition of reciprocity at the level of single links~\cite{kim2023reciprocity}. While this approach can capture different nuances, it is computationally expensive and less straightforward to interpret, as it provides a continuous value instead of a simple yes-or-no answer to whether an interaction is reciprocated.

\begin{figure*}
    \centering
    \includegraphics[width=\linewidth]{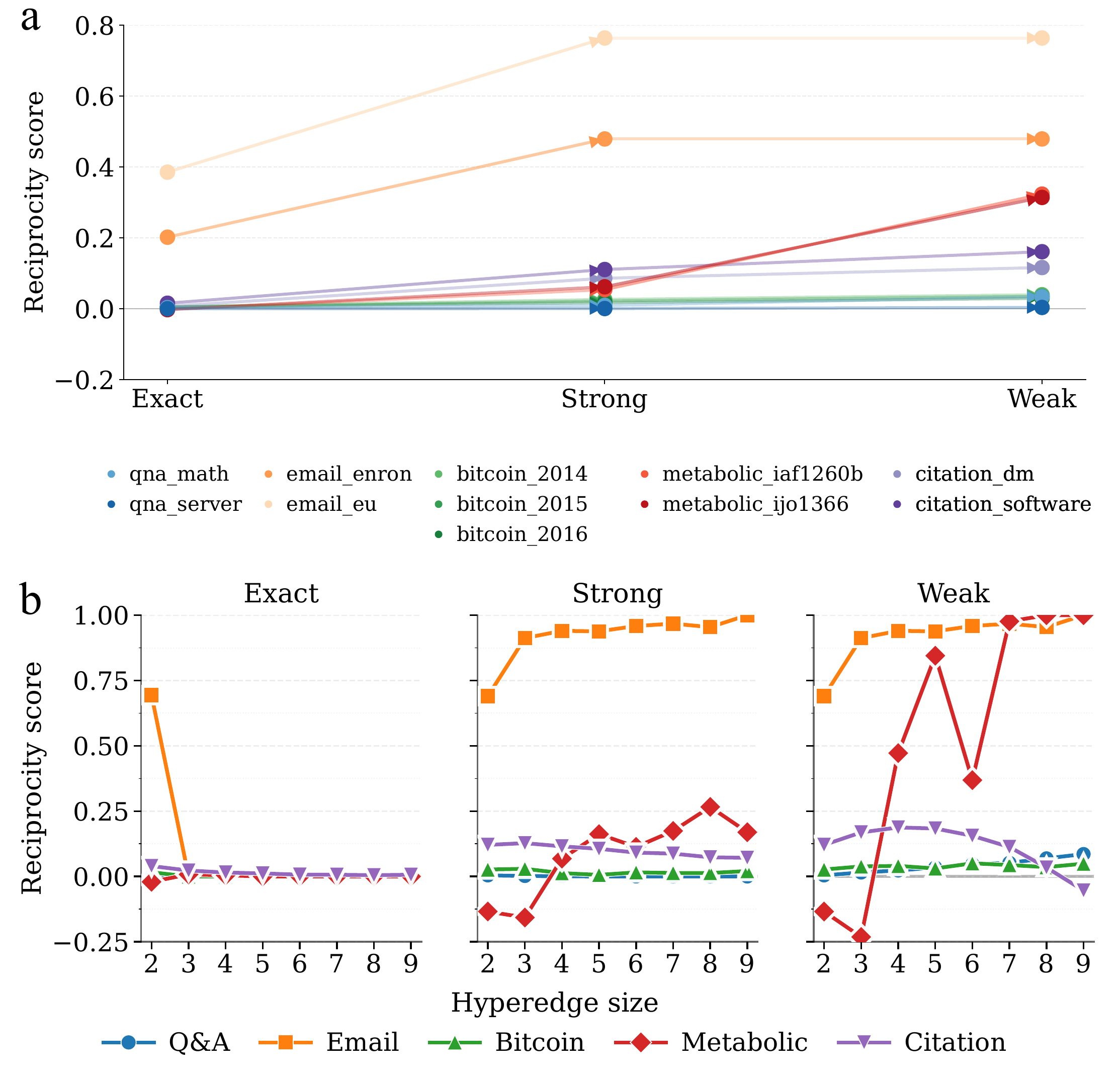}
    \caption{\textbf{Higher-order reciprocity in real-world hypergraphs.} a) Reciprocity score across datasets and reciprocity definitions. Each column corresponds to a distinct notion of higher-order reciprocity, thereby inducing a ranking of the datasets based on their scores. Datasets from the same domains share the same color. Arrows link the datasets across different definitions.
    b) Reciprocity score disaggregated by hyperedge size for each different notion of reciprocity. Trends should be interpreted relative to the null (zero line). To simplify the plots, we aggregate systems from the same domain.}
    \label{fig:reciprocity_res}
\end{figure*}

Here, we introduce three simple and computationally efficient measures for higher-order reciprocity in directed hypergraphs, capturing different aspects of mutual interactions: 

\begin{itemize}
    \item \textbf{Exact reciprocity} occurs when an interaction represented by a hyperedge with a source set \( h \) and a target set \( t \) is precisely mirrored by another interaction with the source and target sets reversed. Formally, two hyperedges \( e_1 = (h_1, t_1) \) and \( e_2 = (h_2, t_2) \) are exactly reciprocated if and only if $h_1 = t_2 \text{ and } t_1 = h_2$. This is the strictest form of reciprocity.
    \item \textbf{Strong reciprocity} relaxes the previous requirement and allows source and target sets to be reversed through a combination of hyperedges, instead of requiring a direct reversal with a single opposite one. Formally, a hyperedge \( e = (h, t) \) is \textit{strongly reciprocated} if there exists a set of hyperedges \( \{ e_1, e_2, \ldots, e_k \} \) such that the union of the target sets of \( e_1, \ldots, e_k \) is a superset of the source set \( h \), and the union of the source sets of \( e_1, \ldots, e_k \) is a superset of the target set \( t \).
    \item \textbf{Weak reciprocity} represents the most relaxed form of reciprocity and requires only that at least one node from the target set of a hyperedge appears in the source set of another, and vice versa. Formally, a hyperedge \( e = (h, t) \) is \textit{weakly reciprocated} if there exists another hyperedge \( e' = (h', t') \) such that $h \cap t' \neq \emptyset \text{ and } t \cap h' \neq \emptyset$.
\end{itemize}

We summarize our definitions of reciprocity for directed hypergraphs in Figure~\ref{fig:reciprocity_def}. More information about the algorithmic aspects of such measures is available in the Methods section. 

After introducing these definitions, a natural first question is which systems exhibit the highest and lowest levels of reciprocity and how the ranking of systems based on reciprocity changes across different definitions. We address this in Figure~\ref{fig:reciprocity_res}a, which shows the normalized ratio of reciprocated hyperedges $\rho_{\mathrm{NM}}$ (reciprocity score) for each system across varying notions of reciprocity. The reciprocity score induces rankings of the systems, allowing us to observe which systems exhibit stronger tendencies toward mutual exchange of information. By definition, the score tends to increase for each system as we move from stricter definitions of reciprocity (exact) to more relaxed ones (weak). We observe that systems from the same domain tend to show similar levels of reciprocity across definitions, indicating that functional similarities within domains may drive comparable reciprocity patterns. \Email datasets exhibit the highest levels of reciprocity, while \bitcoin datasets consistently show the lowest. Interestingly, while the ranking of systems remains largely stable with varying definitions, the relative distances between the datasets change. For instance, exact reciprocity mostly characterizes \Email datasets, which are positioned far from the other datasets, clustering distinctly at the top of the scale. Strong reciprocity induces three clear clusters of datasets based on their scores: \Email datasets rank the highest by a large margin, while \bitcoin datasets occupy the very low end. In the case of weak reciprocity, the datasets begin to separate along domain lines, spanning the entire spectrum of reciprocity scores. Notably, we observe a reduction in the distance between \Email and datasets from metabolic and citation domains, suggesting a convergence in reciprocity levels as the definition becomes more relaxed. Overall, these patterns highlight how the choice of measure can influence the perceived level of reciprocity within different systems. By analyzing how the score evolves across definitions, we gain a more precise understanding of the extent of mutual exchange within each system, from the high reciprocity observed in \Email datasets, where high mutual exchange is clear, to the lower reciprocity in \bitcoin datasets, where reciprocal connections are minimal across all definitions, and to the metabolic datasets, which emerge with high reciprocity under weaker definitions.

A related question is how the size of hyperedges influences the levels of reciprocity. We address this in Figure~\ref{fig:reciprocity_res}b, which reports the reciprocity score as a function of interaction size for the three notions, with deviations interpreted relative to the null baseline. For the \Email datasets, $\rho_{\mathrm{NM}}$ displays consistent positive offset for strong and weak reciprocity across all admissible sizes, whereas exact reciprocity exhibits a pronounced signal only at size~2 and is negligible at larger sizes. This suggests that excess reciprocal structure relative to the null is present for most interaction sizes, while exact "mirror" reversals remain largely a dyadic feature. In the \metabolic datasets, exact reciprocity remains near zero across sizes, with strong and weak deviations showing initial negative values for small sizes, followed by a substantial increase at higher sizes that is more pronounced for the weak definition. Thus, small-set interactions are under-reciprocated relative to the null, whereas large-set interactions become increasingly over-represented. In the \Citation datasets, exact reciprocity is modestly positive at small sizes. Strong reciprocity displays a higher, stable excess across sizes, while weak reciprocity reaches its highest value for the smallest sets and diminishes with size. Citation reciprocity above the null is thus dominated by weak reciprocity in small interactions, with strong reciprocity contributing a more size-invariant offset. For the \bitcoin datasets, scores for all definitions cluster near zero, with a slight increase from exact to strong to weak, and minimal size dependence. This points to weak, size-stable excess reciprocal structure. Finally, the \qna datasets track the \bitcoin pattern, with modest positive offsets that increase from exact to strong to weak definitions, and little systematic variation as size increases.

These findings suggest that weaker notions of reciprocity are valuable in providing insights into the overall reciprocity of systems with larger interactions, capturing a multifaceted view of how reciprocity operates at different strengths and interaction sizes.

\subsection*{Motif analysis in directed hypergraphs}

\begin{figure}
    \centering
    \includegraphics[width=\linewidth]{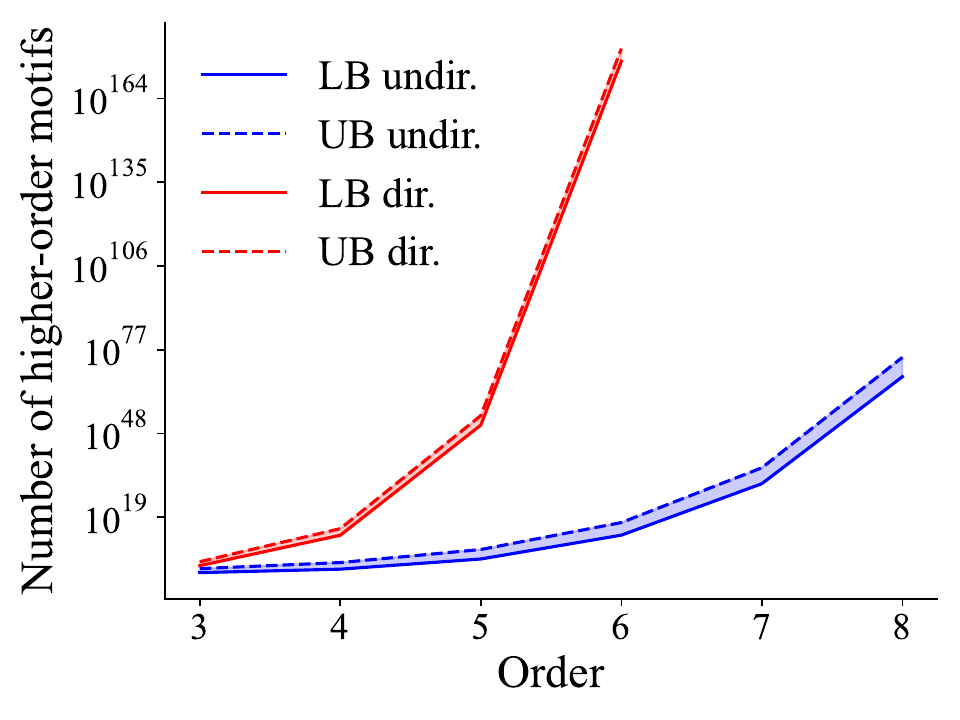}
    \caption{\textbf{Combinatorics of directed higher-order motifs.} Upper (dashed lines) and lower (solid lines) bounds on the number of higher-order motifs as a function of their order. Blue lines refer to undirected motifs on hypergraphs, red lines refer to the directed case.}
    \label{fig:bounds}
\end{figure}
Motif analysis involves counting the frequency of patterns of interactions in connected subgraphs of a given number of nodes. This framework was first introduced by Milo et al.~\cite{milo2002network} to extract the fundamental functional units of complex systems~\cite{milo2004superfamilies}. Recently, motif analysis has been extended to undirected hypergraphs to capture patterns of interactions with arbitrary size~\cite{lotito2022higher}. Here, we extend such analysis to consider also the direction of the hyperedges involved in the patterns.

\begin{figure*}
    \centering
    \includegraphics[width=\linewidth]{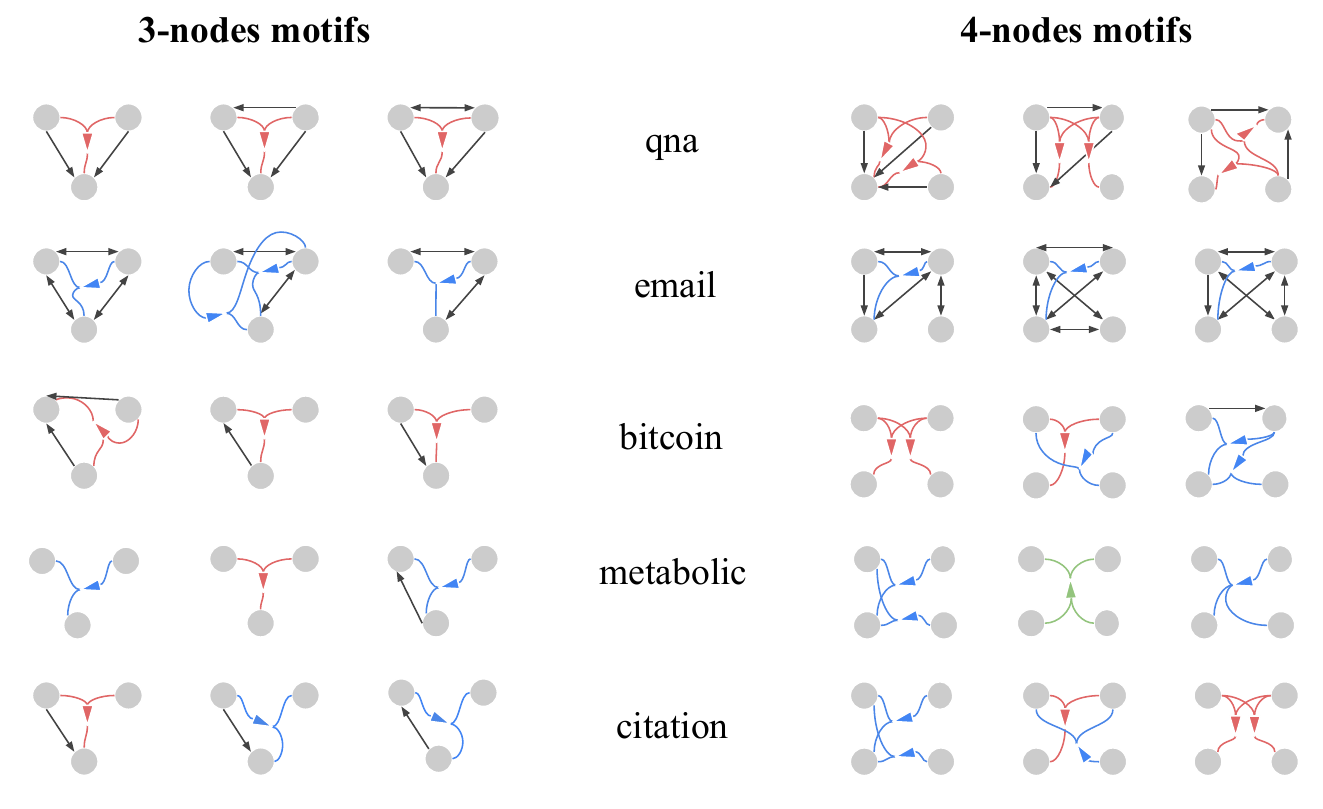}
    \caption{\textbf{Directed higher-order motifs in real-world hypergraphs.} The three most representative directed higher-order motifs of orders three and four from each system. The color of a group interaction encodes its type: one-to-one (black), one-to-many (blue), many-to-one (red), and many-to-many (green). We group statistics of systems within the same domain.}
    \label{fig:motifs}
\end{figure*}

First, it is interesting to study the combinatorics of the patterns of directed subhypergraphs. There is no simple closed-form formula for counting the number of possible directed higher-order motifs as a function of their order \( n \), i.e., the number of nodes in the patterns. We can estimate the number of non-isomorphic connected directed hypergraphs in a way similar to~\cite{lotito2022higher}. Given a set of \( n \) nodes, the number of possible directed hyperedges is \( 3^n - 2 \cdot \sum_{k=1}^{n} \binom{n}{k} - 1 = 3^n - 2 \cdot 2^n + 1 \). This expression counts the ways to partition the \( n \) nodes into three disjoint sets: source, target and empty set. We subtract the invalid combinations with empty source or target sets. Given $n$ nodes, we ensure connectivity by selecting a chain of \( n-1 \) hyperedges and including them in the hypergraph, leaving us with \( 3^n - 2 \cdot 2^n - n + 2 \) remaining possible hyperedges. For each remaining hyperedge, we decide whether to include it or not, resulting in \( 2^{3^n - 2 \cdot 2^n - n + 2} \) total hypergraphs. Since we are interested in non-isomorphic hypergraphs, we divide this number by \( n! \), the number of ways to label the vertices, providing the lower bound \( \frac{2^{3^n - 2 \cdot 2^n - n + 2}}{n!} \). If we ignore the constraints of non-isomorphism and connectivity, we count the number of possible labeled hypergraphs. Since each of the \( 3^n - 2 \cdot 2^n + 1 \) possible hyperedges can either be included or excluded, the total number of labeled hypergraphs is at most \( 2^{3^n - 2 \cdot 2^n + 1} \). Figure~\ref{fig:bounds} shows the upper and lower bounds on the growth of possible sub-hypergraph patterns as a function of the number of nodes (order) for both the undirected and directed cases. The estimated number of patterns grows super-exponentially, even in the undirected case. In the directed case, the growth is even faster due to the need to consider all possible subdivisions into source and target sets. 

To perform motif analysis on real-world directed hypergraphs, we propose an exact algorithm to count the frequency of all connected sub-hypergraph patterns and associate each pattern with a z-score that quantifies its over- or under-representation relative to our proposed configuration model (see Methods).

Given the intractability of the problem for large sub-hypergraphs, we limit our study of empirical data to patterns involving three and four nodes. Moreover, we focus on patterns that include at least one group interaction. Because the theoretical pattern space is still very large, while the empirical signal is sparse and concentrated (see Supplementary Note~3), we restrict our analysis to the positive tail of the motif z-score distributions (see Supplementary Note~3) to highlight structural differences across datasets. In particular, Figure~\ref{fig:motifs} shows the most over-represented patterns of directed higher-order interactions with three and four nodes across different domains. Each domain reveals distinct motifs, characterized by different directed hyperedge types, sizes, densities and patterns of reciprocity. In terms of hyperedges types, \Email and \qna involve abundant patterns with only many-to-one and one-to-many interactions. Other datasets display more diverse patterns, including combinations of one-to-many, many-to-one, and many-to-many interactions (the latter is possible only in motifs with four nodes). Traditional one-to-one interactions are commonly part of abundant patterns in all datasets. The number of interactions in abundant sub-hypergraphs is small in the \bitcoin, \metabolic and \Citation domains, often involving just one or two hyperedges. In contrast, the \Email and \qna domains tend to be richer in interactions. This observation is reversed when considering the average size of interactions. The relation between the number and the average size of interactions aligns with previous studies on undirected higher-order motifs~\cite{lotito2022higher}. A common pattern in many datasets is the coexistence of group interactions alongside lower-order interactions within the same set of nodes, a phenomenon often referred to as nestedness~\cite{lotito2022higher} or simpliciality~\cite{landry2024simpliciality}.
Interestingly, considering the direction of such interactions reveals that they seem to play a role in increasing the overall reciprocity of the patterns, suggesting the existence of a feedback mechanism. This is particularly evident in \Email data. In addition to reciprocity, the direction of lower-order interactions in abundant patterns suggests a reinforcing mechanism where subsets of source and target nodes interact at multiple interaction sizes. These observations are closely connected with the insights discussed in the previous sections about frequent co-senders and co-receivers nodes and higher-order reciprocity.

\section*{Discussion}
Hypergraphs extend traditional network representations by allowing hyperedges to connect multiple nodes simultaneously, enabling the encoding of group interactions ubiquitous in many relational systems. Directed hypergraphs further enhance our modelling abilities by accounting for directionality in group interactions, distinguishing between source and target sets for each hyperedge. This versatile framework can accurately model a range of diverse real-world systems and interactions, including financial transactions, email exchanges, and metabolic reactions. 

In this work, we proposed new measures and tools to analyze the structural organization of directed hypergraphs at their microscale. First, we analyzed hyperedge signature vectors to identify the abundance of each hyperedge structure across datasets and identified classes of systems sharing similar higher-order connectivity patterns. Second, we analyzed the excess overlap among source and target sets for each node in each system. The resulting distributions suggest that different domains may follow distinct organizational principles, ranging from redundant to more diverse interaction patterns. Then, we introduced three distinct types of higher-order reciprocity measures: exact, strong, and weak reciprocity. Each definition offers a different perspective on how group interactions can be reciprocated, ranging from strict to more relaxed forms of reciprocal influence, and can be computed efficiently, making it suitable also for the analysis of very large systems. We showed that all systems exhibit reciprocity in broad terms, though different domains are associated with specific patterns and sensitivity to specific reciprocity measures. Lastly, we extended the notion of motifs to directed hypergraphs, capturing recurring patterns of directed interactions. Motif analysis revealed frequent microscale structures and highlighted common organizational principles playing a role in the function and behavior of systems, such as the existence of reinforcing or feedback mechanisms among dyadic and non-dyadic interactions in groups. 

Taken together, by considering the nuances related to the directionality of interactions in directed hypergraphs, our research provides a framework to understand higher-order connectivity in directed complex systems, opening up a wide range of potential applications in diverse fields such as social network analysis, biology, and finance. For instance, the study of multi-party financial transactions as directed higher-order structures may capture more complex patterns of fraudulent activity than traditional graph-based models~\cite{akoglu2015graph}. Similarly, directed hypergraphs may enhance the accuracy of existing frameworks in identifying and predicting important genes based on genomic expression relations~\cite{feng2021hypergraph}. As scalability is a pressing issue in hypergraph algorithms, future work may explore more advanced techniques for detecting motifs in large-scale directed hypergraphs. These include sampling methods already proposed for undirected higher-order motifs~\cite{lotito2024exact} or the use of parallel algorithms, which may achieve significant speed-ups beyond the implementation of our current Python library, thereby enabling analysis of motifs larger than four nodes. Another interesting venue for further studies is related to the study of reciprocity in time-evolving hypergraphs, since it can affect mechanisms of group formation~\cite{cencetti2021temporal, iacopini2024temporal} and inform the efficient seeding of information~\cite{mancastroppa2023hyper, genetti2024influence}.
All in all, our work reveals new structural principles behind the organization of real-world systems, shedding light on the complex interplay between structural patterns and functionality in directed complex systems.

\begin{widetext}

\section*{Methods}

\subsection*{Hyperedge signature vector construction}

For each dataset, we construct a \textit{hyperedge signature vector} \( \mathbf{v} \), where each element represents the count of hyperedges with a specific combination of source set size \( s \) and target set size \( t \) in the hypergraph. The vector \( \mathbf{v} \) captures the distribution of hyperedges based on the sizes of their source and target sets, providing a profile of the hypergraph structure.

Formally, we define the vector \( \mathbf{v} \) as follows:
\[
\mathbf{v} = \left( v_{1,2}, \ldots, v_{1,K-1}, v_{2,1}, \ldots, v_{2,K-2}, \ldots, v_{K-1,1} \right)
\]
where \( K \) represents the maximum hyperedge size considered, and each \( v_{h,t} \) counts the number of hyperedges with a specific source size \( h \) and target size \( t \).

\subsection*{Microcanonical set-swap configuration model for directed hypergraphs}
We generate randomized counterparts of directed hypergraphs using a microcanonical set-swap configuration model. This approach is similar to the one proposed by Preti et al.~\cite{preti2024higher}, and more recently by Kraakman et al.~\cite{kraakman2025hypercurveball}. More broadly, our null model fits within the family of configuration- and entropy-based random hypergraph ensembles that preserve degree and edge-size sequences~\cite{chodrow2020configuration,barthelemy2022class,nakajima2021randomizing,saracco2025entropy}, which have been used in applications such as extracting statistically validated higher-order interactions~\cite{musciotto2021detecting} and motifs~\cite{lotito2022higher}. In our framework, each directed hyperedge \(e\in E\) is specified by two disjoint node sets: a source set \(s(e)\subseteq V\) and a target set \(t(e)\subseteq V\), with \(s(e)\cap t(e)=\varnothing\). The microcanonical ensemble comprises all simple directed hypergraphs that match the observed node out-/in-degree sequences (counts of appearances in \(s(\cdot)\) and \(t(\cdot)\), respectively) and preserve every hyperedge’s cardinalities \(|s(e)|\) and \(|t(e)|\). We forbid duplicated hyperedges.

Starting from the observed \(H\), we repeatedly attempt \emph{set-swaps} on one side at a time. In a source-side attempt, we select two distinct hyperedges \(e_i\neq e_j\) uniformly and sample \(u\in s(e_i)\) and \(v\in s(e_j)\). We propose
\[
s'(e_i)=(s(e_i)\setminus\{u\})\cup\{v\},\qquad
s'(e_j)=(s(e_j)\setminus\{v\})\cup\{u\},
\]
leaving \(t(e_i)\) and \(t(e_j)\) unchanged. The move is accepted only if it respects set semantics on the chosen side (no duplicates), preserves within-edge disjointness (\(s(e)\cap t(e)=\varnothing\)), and does not create a duplicate \((s'(e),t(e))\). Target-side attempts are defined symmetrically. Accepted swaps conserve each node’s in/out degree and all \(|s(e)|,|t(e)|\). Rejected proposals leave the state unchanged.

\subsection*{Hyperedge overlap}
For each node, we quantify its hyperedge overlap separately for its participation in target sets (in-hyperedges) and source sets (out-hyperedges). For in-hyperedge overlap, we consider all hyperedges \(e\) for which the node is in the target set \(t(e)\). Let \(\mathcal{E}_{\text{in}}\) be the collection of these hyperedges. We define the in-hyperedge overlap as
\[
O_{\text{in}} = \frac{\sum_{e \in \mathcal{E}_{\text{in}}} |t(e)|}{|\mathcal{E}_{\text{in}}| \left|\bigcup_{e \in \mathcal{E}_{\text{in}}} t(e)\right|},
\]
where \(|t(e)|\) denotes the number of nodes in the target set of hyperedge \(e\).

Similarly, for out-hyperedge overlap, we consider only hyperedges \(e\) in which the node is a source (i.e., belongs to \(s(e)\)). Let \(\mathcal{E}_{\text{out}}\) denote this collection, and define
\[
O_{\text{out}} = \frac{\sum_{e \in \mathcal{E}_{\text{out}}} |s(e)|}{|\mathcal{E}_{\text{out}}| \left|\bigcup_{e \in \mathcal{E}_{\text{out}}} s(e)\right|},
\]
with \(|s(e)|\) being the size of the source set of hyperedge \(e\). These metrics yield a value of 1 when all corresponding hyperedges share an identical set of nodes (i.e., maximal overlap), and decrease as the sets become more diverse.

To assess statistical significance, we compare the observed overlaps to those computed on an ensemble of randomized networks that preserve key structural properties (e.g., node degrees and hyperedge sizes). For each node \(i\), we compute a z-score that standardizes the observed overlap relative to the null ensemble:
\begin{equation*}
z_O(i) = \frac{O_i^{\mathrm{obs}} - \langle O_i^{\mathrm{rand}} \rangle}{\sqrt{\operatorname{Var}\!\left(O_i^{\mathrm{rand}}\right)}}\,.
\end{equation*}
We compute this for both \(O_{\text{sources}}\) and \(O_{\text{targets}}\).

\subsection*{Algorithms for measuring reciprocity in directed hypergraphs}
Below, we outline our proposed algorithms for efficiently measuring reciprocity in directed hypergraphs.

\begin{itemize}
    \item \textbf{Exact reciprocity.} Each hyperedge $e = (s, t)$ is stored in a hash-based dictionary, and for each hyperedge, we search for a reverse hyperedge $e' = (t, s)$. Since each lookup takes constant time, the overall complexity is $O(m)$, where $m$ is the number of hyperedges.
    \item \textbf{Strong reciprocity.} For each hyperedge $e = (s, t)$, we maintain a reachability dictionary that tracks which nodes in the target set $t$ can reach other nodes via multiple hyperedges. We then check whether the source set $s$ is fully covered by the accumulated reachable nodes from the target set $t$. This involves iterating over each hyperedge, for each target node, accumulating the reachable nodes and then checking if the source set is a subset of this accumulated set. Computing the union of reachable nodes is $O(s \cdot t)$, where $s$ is the maximum size of source sets and $t$ is the maximum size of target sets. This operation is repeated for all hyperedges, leading to a total complexity of $O(m \cdot s \cdot t)$.
    \item \textbf{Weak reciprocity.} First, we construct a dictionary to store all directed node pairs between the source and target sets of each hyperedge. Then, for each hyperedge, we check whether any of its target nodes are linked back to the source nodes via reverse connections in the dictionary. The computational complexity is dominated by the first operation, which is $O(m \cdot s \cdot t)$, where $s$ is the maximum size of the source sets and $t$ is the maximum size of the target sets across all hyperedges.
\end{itemize}

In practice, executing these algorithms on the real-world datasets used in our experiments requires only a few minutes for all datasets combined, demonstrating the computational efficiency of the proposed methods.

\subsection*{Algorithms for motif analysis in directed hypergraphs}
In order to design efficient algorithms for mining directed higher-order motifs, we extend prior ideas developed for the same problem in undirected hypergraphs~\cite{lotito2024exact}. Our algorithms are efficient enough to count motifs of size $3$ and $4$ in datasets of reasonable size (comparable to those used in our experiments). However, scaling to larger datasets and motifs of larger size would require more sophisticated approaches, such as sampling algorithms~\cite{lotito2024exact}, which we leave for future work. Further details on the execution times of the algorithms for mining motifs of orders $3$ and $4$ can be found in Supplementary Notes $2$.

The algorithm for mining motifs (involving at least one group interaction) of order $3$ begins by iterating through each hyperedge in the hypergraph that contains exactly three vertices. For each such hyperedge, it identifies all possible subsets of vertices and checks whether one or more subsets form valid directed hyperedges in the hypergraph. Valid subsets, along with the original hyperedge, define the motif structure involving those three vertices. To ensure consistency in motif identification, the algorithm generates a canonical form of the motif by lexicographical ordering its vertices and edges, which can be computed by sorting the $n!$ possible relabels. This canonical representation allows motifs with the same structural pattern to be compared and counted, even if they differ in their vertex labels. Each canonical form of motifs is stored in a frequency hash map. If the motif has not been encountered before, it is added to the map; if it has, its frequency count is incremented. In the end, the algorithm outputs a distribution of the various motif structures of order $3$. This algorithm operates in linear time with respect to the number of hyperedges of order $3$. Specifically, its computational complexity is \( O(m_3) \), where \( m_3 \) is the number of hyperedges involving exactly three vertices. Each motif construction and comparison is performed in constant time due to the fixed size of the motifs. For more details, refer to the pseudocode in Supplementary Note $2$. The algorithm for mining motifs of order $4$ follows a similar approach. First, it iterates over all hyperedges of size $4$, counting the motifs involving exactly these $4$ nodes. Unlike the previous algorithm, it then iterates over all hyperedges of size $3$, performing an additional neighborhood exploration step to identify the fourth node involved in the motif. Each neighboring node is considered during this process. Once the 4 nodes are identified, the algorithm constructs the motif as before. The pseudocode for this algorithm is provided in Supplementary Note $2$.

\subsection*{Statistical significance of motifs}
To distinguish meaningful, non-random interaction patterns from those that may occur by chance, we use a configuration model as a null model to evaluate the statistical significance of the interaction patterns after computing their frequency in our directed hypergraphs. The configuration model generates randomized versions of the original hypergraph while preserving key properties, such as the in-degree and out-degree sequences, as well as the source and target sizes of the hyperedges~\cite{preti2024higher}. By comparing the observed frequencies with those found in the randomized networks, we can identify significantly over-represented motifs.

In particular, each motif \( i \) is associated with a standardized score \( z_M(i) \), which quantifies how many standard deviations the observed frequency of the motif differs from its expected value under the configuration model~\cite{milo2002network, lotito2022higher}. A larger absolute value of \( z_M(i) \) indicates a stronger statistical deviation from the null expectation, meaning the motif is significantly over- or under-represented compared to random networks. The score is defined as:

\begin{equation*}
z_M(i) = \frac{N_i^{\mathrm{obs}} - \langle N_i^{\mathrm{rand}} \rangle}{\sqrt{\operatorname{Var}\!\left(N_i^{\mathrm{rand}}\right)}}\,.
\end{equation*}

Here, \( N_i^{\mathrm{obs}} \) denotes the observed frequency of motif \( i \) in the empirical hypergraph, and \( \langle N_i^{\mathrm{rand}} \rangle \) and \( \operatorname{Var}\!\left(N_i^{\mathrm{rand}}\right) \) are the mean and variance of motif \( i \)’s frequency across random hypergraphs generated by the configuration model. Following~\cite{milo2004superfamilies}, we generate 100 random samples from the configuration model for each hypergraph to estimate the mean and variance of the motif frequencies.

\section*{Code availability}
The tools for the analysis of directed hypergraphs presented in this work are available as part of Hypergraphx (HGX)~\cite{lotito2023hypergraphx}.

\section*{Data availability} 
Data~\cite{kim2023reciprocity} is publicly available and also easily accessible through HGX~\cite{lotito2023hypergraphx}.

\section*{Acknowledgements}
F.B. acknowledges support from the Air Force Office of Scientific Research under award number FA8655-22-1-7025. F.B. acknowledges support from the Austrian Science Fund (FWF) through project 10.55776/PAT1052824 and project 10.55776/PAT1652425. A.M. acknowledges support from the European Union through Horizon Europe CLOUDSTARS project (101086248).

\end{widetext}

\bibliography{biblio}

@article{civilini2024explosive,
  title={Explosive cooperation in social dilemmas on higher-order networks},
  author={Civilini, Andrea and Sadekar, Onkar and Battiston, Federico and G{\'o}mez-Garde{\~n}es, Jes{\'u}s and Latora, Vito},
  journal={Physical Review Letters},
  volume={132},
  number={16},
  pages={167401},
  year={2024},
  publisher={APS}
}

@article{preti2024higher,
  title={Higher-Order Null Models as a Lens for Social Systems},
  author={Preti, Giulia and Fazzone, Adriano and Petri, Giovanni and De Francisci Morales, Gianmarco},
  journal={Physical Review X},
  volume={14},
  number={3},
  pages={031032},
  year={2024},
  publisher={APS}
}

@article{moon2023four,
  title={Four-set hypergraphlets for characterization of directed hypergraphs},
  author={Moon, Heechan and Kim, Hyunju and Kim, Sunwoo and Shin, Kijung},
  journal={arXiv preprint arXiv:2311.14289},
  year={2023}
}

@article{kim2023reciprocity,
  title={Reciprocity in directed hypergraphs: measures, findings, and generators},
  author={Kim, Sunwoo and Choe, Minyoung and Yoo, Jaemin and Shin, Kijung},
  journal={Data Mining and Knowledge Discovery},
  volume={37},
  number={6},
  pages={2330--2388},
  year={2023},
  publisher={Springer}
}

@article{gallo1993directed,
title = {Directed hypergraphs and applications},
journal = {Discrete Applied Mathematics},
volume = {42},
number = {2},
pages = {177-201},
year = {1993},
issn = {0166-218X},
doi = {https://doi.org/10.1016/0166-218X(93)90045-P},
url = {https://www.sciencedirect.com/science/article/pii/0166218X9390045P},
author = {Giorgio Gallo and Giustino Longo and Stefano Pallottino and Sang Nguyen}
}

@article {milo2004superfamilies,
	author = {Milo, Ron and Itzkovitz, Shalev and Kashtan, Nadav and Levitt, Reuven and Shen-Orr, Shai and Ayzenshtat, Inbal and Sheffer, Michal and Alon, Uri},
	title = {Superfamilies of Evolved and Designed Networks},
	volume = {303},
	number = {5663},
	pages = {1538--1542},
	year = {2004},
	doi = {10.1126/science.1089167},
	publisher = {American Association for the Advancement of Science},
	issn = {0036-8075},
	journal = {Science}
}

@article{milo2002network,
  title = {Network Motifs: Simple Building Blocks of Complex Networks},
  author = {Milo, Ron and {Shen-Orr}, Shai and Itzkovitz, Shalev and Kashtan, Nadav and Chklovskii, Dmitri and Alon, Uri},
  year = {2002},
  volume = {298},
  pages = {824--827},
  publisher = {{American Association for the Advancement of Science}},
  journal = {Science},
  number = {5594}
}

@article{lotito2023hypergraphx,
  title={Hypergraphx: a library for higher-order network analysis},
  author={Lotito, Quintino Francesco and Contisciani, Martina and De Bacco, Caterina and Di Gaetano, Leonardo and Gallo, Luca and Montresor, Alberto and Musciotto, Federico and Ruggeri, Nicol{\`o} and Battiston, Federico},
  journal={Journal of Complex Networks},
  volume={11},
  number={3},
  pages={cnad019},
  year={2023},
  publisher={Oxford University Press}
}

@article{battiston2020networks,
  title={Networks beyond pairwise interactions: structure and dynamics},
  author={Battiston, Federico and Cencetti, Giulia and Iacopini, Iacopo and Latora, Vito and Lucas, Maxime and Patania, Alice and Young, Jean-Gabriel and Petri, Giovanni},
  journal={Physics Reports},
  volume={874},
  pages={1--92},
  year={2020},
  publisher={Elsevier}
}

@article{battiston2021physics,
  title={The physics of higher-order interactions in complex systems},
  author={Battiston, Federico and Amico, Enrico and Barrat, Alain and Bianconi, Ginestra and Ferraz de Arruda, Guilherme and Franceschiello, Benedetta and Iacopini, Iacopo and K{\'e}fi, Sonia and Latora, Vito and Moreno, Yamir and others},
  journal={Nature Physics},
  volume={17},
  number={10},
  pages={1093--1098},
  year={2021},
  publisher={Nature Publishing Group}
}

@article{benson2019three,
  title={Three hypergraph eigenvector centralities},
  author={Benson, Austin R},
  journal={SIAM Journal on Mathematics of Data Science},
  volume={1},
  number={2},
  pages={293--312},
  year={2019},
  publisher={SIAM}
}

@book{berge1973graphs,
  title={Graphs and hypergraphs},
  author={Berge, Claude},
  year={1973},
  publisher={North-Holland Pub. Co.}
}

@article{boccaletti2006complex,
  title={Complex networks: Structure and dynamics},
  author={Boccaletti, Stefano and Latora, Vito and Moreno, Yamir and Chavez, Martin and Hwang, D-U},
  journal={Physics Reports},
  volume={424},
  number={4-5},
  pages={175--308},
  year={2006},
  publisher={Elsevier}
}

@article{carletti2020random,
  title={Random walks on hypergraphs},
  author={Carletti, Timoteo and Battiston, Federico and Cencetti, Giulia and Fanelli, Duccio},
  journal={Physical Review E},
  volume={101},
  number={2},
  pages={022308},
  year={2020},
  publisher={APS}
}

@article{cencetti2021temporal,
  title={Temporal properties of higher-order interactions in social networks},
  author={Cencetti, Giulia and Battiston, Federico and Lepri, Bruno and Karsai, M{\'a}rton},
  journal={Scientific Reports},
  volume={11},
  number={1},
  pages={1--10},
  year={2021},
  publisher={Nature Publishing Group}
}

@article{chodrow2020configuration,
  title={Configuration models of random hypergraphs},
  author={Chodrow, Philip S},
  journal={Journal of Complex Networks},
  volume={8},
  number={3},
  pages={cnaa018},
  year={2020},
  publisher={Oxford University Press}
}

@article{chowdhary2021simplicial,
  title={Simplicial contagion in temporal higher-order networks},
  author={Chowdhary, Sandeep and Kumar, Aanjaneya and Cencetti, Giulia and Iacopini, Iacopo and Battiston, Federico},
  journal={Journal of Physics: Complexity},
  volume={2},
  number={3},
  pages={035019},
  year={2021},
  publisher={IOP Publishing}
}

@article{contisciani2022inference,
  title={Inference of hyperedges and overlapping communities in hypergraphs},
  author={Contisciani, Martina and Battiston, Federico and De Bacco, Caterina},
  journal={Nature communications},
  volume={13},
  number={1},
  pages={1--10},
  year={2022},
  publisher={Nature Publishing Group}
}

@article{eriksson2021choosing,
  title={How choosing random-walk model and network representation matters for flow-based community detection in hypergraphs},
  author={Eriksson, Anton and Edler, Daniel and Rojas, Alexis and de Domenico, Manlio and Rosvall, Martin},
  journal={Communications Physics},
  volume={4},
  number={1},
  pages={1--12},
  year={2021},
  publisher={Nature Publishing Group}
}

@article{santoro2023higher,
  title={Higher-order connectomics of human brain function reveals local topological signatures of task decoding, individual identification, and behavior},
  author={Santoro, Andrea and Battiston, Federico and Lucas, Maxime and Petri, GIovanni and Amico, Enrico},
  journal={bioRxiv},
  pages={2023--12},
  year={2023},
  publisher={Cold Spring Harbor Laboratory}
}

@article{gambuzza2021stability,
  title={Stability of synchronization in simplicial complexes},
  author={Gambuzza, Lucia Valentina and Di Patti, Francesca and Gallo, Luca and Lepri, Stefano and Romance, Miguel and Criado, Regino and Frasca, Mattia and Latora, Vito and Boccaletti, Stefano},
  journal={Nature Communications},
  volume={12},
  number={1},
  pages={1--13},
  year={2021},
  publisher={Nature Publishing Group}
}

@article{grilli2017higher,
  title={Higher-order interactions stabilize dynamics in competitive network models},
  author={Grilli, Jacopo and Barab{\'a}s, Gy{\"o}rgy and Michalska-Smith, Matthew J and Allesina, Stefano},
  journal={Nature},
  volume={548},
  number={7666},
  pages={210--213},
  year={2017},
  publisher={Nature Publishing Group}
}

@article{iacopini2019simplicial,
  title={Simplicial models of social contagion},
  author={Iacopini, Iacopo and Petri, Giovanni and Barrat, Alain and Latora, Vito},
  journal={Nature Communications},
  volume={10},
  number={1},
  pages={1--9},
  year={2019},
  publisher={Nature Publishing Group}
}

@article{digaetano2024percolation,
  title={Percolation and topological properties of temporal higher-order networks},
  author={Di Gaetano, Leonardo and Battiston, Federico and Starnini, Michele},
  journal={Physical Review Letters},
  volume={132},
  number={3},
  pages={037401},
  year={2024},
  publisher={APS}
}

@article{gallo2024higher,
  title={Higher-order correlations reveal complex memory in temporal hypergraphs},
  author={Gallo, Luca and Lacasa, Lucas and Latora, Vito and Battiston, Federico},
  journal={Nature Communications},
  volume={15},
  number={1},
  pages={4754},
  year={2024},
  publisher={Nature Publishing Group UK London}
}

@article{feng2021hypergraph,
  title={Hypergraph models of biological networks to identify genes critical to pathogenic viral response},
  author={Feng, Song and Heath, Emily and Jefferson, Brett and Joslyn, Cliff and Kvinge, Henry and Mitchell, Hugh D and Praggastis, Brenda and Eisfeld, Amie J and Sims, Amy C and Thackray, Larissa B and others},
  journal={BMC bioinformatics},
  volume={22},
  number={1},
  pages={287},
  year={2021},
  publisher={Springer}
}

@article{akoglu2015graph,
  title={Graph based anomaly detection and description: a survey},
  author={Akoglu, Leman and Tong, Hanghang and Koutra, Danai},
  journal={Data mining and knowledge discovery},
  volume={29},
  pages={626--688},
  year={2015},
  publisher={Springer}
}

@article{pearcy2014hypergraph,
  title={Hypergraph models of metabolism},
  author={Pearcy, Nicole and Crofts, Jonathan J and Chuzhanova, Nadia},
  journal={International Journal of Biological, Veterinary, Agricultural and Food Engineering},
  volume={8},
  number={8},
  pages={752--756},
  year={2014},
  publisher={World Academy of Science, Engineering and Technology}
}

@article{lotito2022higher,
  author = {Lotito, Quintino Francesco and Musciotto, Federico and Montresor, Alberto and Battiston, Federico},
  doi = {10.1038/s42005-022-00858-7},
  issn = {2399-3650},
  journal = {Communications Physics},
  number = {1},
  pages = {79},
  title = {{Higher-order motif analysis in hypergraphs}},
  url = {https://doi.org/10.1038/s42005-022-00858-7},
  volume = {5},
  year = {2022}
}

@article{lee2020hypergraph,
  title={Hypergraph motifs: concepts, algorithms, and discoveries},
  author={Lee, Geon and Ko, Jihoon and Shin, Kijung},
  journal={Proceedings of the VLDB Endowment},
  volume={13},
  number={12},
  pages={2256--2269},
  year={2020},
  publisher={VLDB Endowment}
}

@article{iacopini2024temporal,
  title={The temporal dynamics of group interactions in higher-order social networks},
  author={Iacopini, Iacopo and Karsai, M{\'a}rton and Barrat, Alain},
  journal={Nature Communications},
  volume={15},
  number={1},
  pages={7391},
  year={2024},
  publisher={Nature Publishing Group UK London}
}

@article{arregui2024patterns,
  title={Patterns in temporal networks with higher-order egocentric structures},
  author={Arregui-Garc{\'\i}a, Beatriz and Longa, Antonio and Lotito, Quintino Francesco and Meloni, Sandro and Cencetti, Giulia},
  journal={Entropy},
  volume={26},
  number={3},
  pages={256},
  year={2024},
  publisher={MDPI}
}

@article{lotito2024exact,
  title={Exact and sampling methods for mining higher-order motifs in large hypergraphs},
  author={Lotito, Quintino Francesco and Musciotto, Federico and Battiston, Federico and Montresor, Alberto},
  journal={Computing},
  volume={106},
  number={2},
  pages={475--494},
  year={2024},
  publisher={Springer}
}

@article{lucas2020multiorder,
  title={Multiorder Laplacian for synchronization in higher-order networks},
  author={Lucas, Maxime and Cencetti, Giulia and Battiston, Federico},
  journal={Physical Review Research},
  volume={2},
  number={3},
  pages={033410},
  year={2020},
  publisher={APS}
}

@article{musciotto2021detecting,
  title={Detecting informative higher-order interactions in statistically validated hypergraphs},
  author={Musciotto, Federico and Battiston, Federico and Mantegna, Rosario N},
  journal={Communications Physics},
  volume={4},
  number={1},
  pages={1--9},
  year={2021},
  publisher={Nature Publishing Group}
}

@article{patania2017shape,
  title={The shape of collaborations},
  author={Patania, Alice and Petri, Giovanni and Vaccarino, Francesco},
  journal={EPJ Data Science},
  volume={6},
  pages={1--16},
  year={2017},
  publisher={Springer}
}

@article{petri2018simplicial,
  title={Simplicial activity driven model},
  author={Petri, Giovanni and Barrat, Alain},
  journal={Physical review letters},
  volume={121},
  number={22},
  pages={228301},
  year={2018},
  publisher={APS}
}

@article{gallo2022synchronization,
  title={Synchronization induced by directed higher-order interactions},
  author={Gallo, Luca and Muolo, Riccardo and Gambuzza, Lucia Valentina and Latora, Vito and Frasca, Mattia and Carletti, Timoteo},
  journal={Communications Physics},
  volume={5},
  number={1},
  pages={263},
  year={2022},
  publisher={Nature Publishing Group UK London}
}

@article{petri2014homological,
  title={Homological scaffolds of brain functional networks},
  author={Petri, Giovanni and Expert, Paul and Turkheimer, Federico and Carhart-Harris, Robin and Nutt, David and Hellyer, Peter J and Vaccarino, Francesco},
  journal={Journal of The Royal Society Interface},
  volume={11},
  number={101},
  pages={20140873},
  year={2014},
  publisher={The Royal Society}
}

@article{ruggeri2023community,
  title={Community detection in large hypergraphs},
  author={Ruggeri, Nicol{\`o} and Contisciani, Martina and Battiston, Federico and De Bacco, Caterina},
  journal={Science Advances},
  volume={9},
  number={28},
  pages={eadg9159},
  year={2023},
  publisher={American Association for the Advancement of Science}
}

@article{schaub2020random,
  title={Random walks on simplicial complexes and the normalized hodge 1-laplacian},
  author={Schaub, Michael T and Benson, Austin R and Horn, Paul and Lippner, Gabor and Jadbabaie, Ali},
  journal={SIAM Review},
  volume={62},
  number={2},
  pages={353--391},
  year={2020},
  publisher={SIAM}
}

@article{tudisco2021node,
  title={Node and edge nonlinear eigenvector centrality for hypergraphs},
  author={Tudisco, Francesco and Higham, Desmond J},
  journal={Communications Physics},
  volume={4},
  number={1},
  pages={1--10},
  year={2021},
  publisher={Nature Publishing Group}
}

@article{cimini2019statistical,
  title={The statistical physics of real-world networks},
  author={Cimini, Giulio and Squartini, Tiziano and Saracco, Fabio and Garlaschelli, Diego and Gabrielli, Andrea and Caldarelli, Guido},
  journal={Nature Reviews Physics},
  volume={1},
  number={1},
  pages={58--71},
  year={2019},
  publisher={Nature Publishing Group UK London}
}

@article{garlaschelli2004patterns,
  title={Patterns of link reciprocity in directed networks},
  author={Garlaschelli, Diego and Loffredo, Maria I},
  journal={Physical review letters},
  volume={93},
  number={26},
  pages={268701},
  year={2004},
  publisher={APS}
}

@article{ghoshal2009random,
  title={Random hypergraphs and their applications},
  author={Ghoshal, Gourab and Zlati{\'c}, Vinko and Caldarelli, Guido},
  journal={Physical Review E},
  volume={79},
  number={6},
  pages={066118},
  year={2009},
  publisher={APS}
}

@book{wasserman1994social,
  title={Social network analysis: Methods and applications},
  author={Wasserman, Stanley and Faust, Katherine and others},
  year={1994},
  publisher={Cambridge university press}
}

@article{zhang2023higher,
  title={Higher-order interactions shape collective dynamics differently in hypergraphs and simplicial complexes},
  author={Zhang, Yuanzhao and Lucas, Maxime and Battiston, Federico},
  journal={Nature Communications},
  volume={14},
  number={1},
  pages={1605},
  year={2023},
  publisher={Nature Publishing Group UK London}
}

@inproceedings{yadati2020nhp,
  title={Nhp: Neural hypergraph link prediction},
  author={Yadati, Naganand and Nitin, Vikram and Nimishakavi, Madhav and Yadav, Prateek and Louis, Anand and Talukdar, Partha},
  booktitle={Proceedings of the 29th ACM international conference on information \& knowledge management},
  pages={1705--1714},
  year={2020}
}

@inproceedings{yadati2021graph,
  title={Graph neural networks for soft semi-supervised learning on hypergraphs},
  author={Yadati, Naganand and Gao, Tingran and Asoodeh, Shahab and Talukdar, Partha and Louis, Anand},
  booktitle={Pacific-Asia Conference on Knowledge Discovery and Data Mining},
  pages={447--458},
  year={2021},
  organization={Springer}
}

@inproceedings{ranshous2017exchange,
  title={Exchange pattern mining in the bitcoin transaction directed hypergraph},
  author={Ranshous, Stephen and Joslyn, Cliff A and Kreyling, Sean and Nowak, Kathleen and Samatova, Nagiza F and West, Curtis L and Winters, Samuel},
  booktitle={Financial Cryptography and Data Security: FC 2017 International Workshops, WAHC, BITCOIN, VOTING, WTSC, and TA, Sliema, Malta, April 7, 2017, Revised Selected Papers 21},
  pages={248--263},
  year={2017},
  organization={Springer}
}

@article{traversa2023robustness,
  title={Robustness and complexity of directed and weighted metabolic hypergraphs},
  author={Traversa, Pietro and Ferraz de Arruda, Guilherme and Vazquez, Alexei and Moreno, Yamir},
  journal={Entropy},
  volume={25},
  number={11},
  pages={1537},
  year={2023},
  publisher={MDPI}
}

@article{jost2019hypergraph,
title = {Hypergraph Laplace operators for chemical reaction networks},
journal = {Advances in Mathematics},
volume = {351},
pages = {870-896},
year = {2019},
issn = {0001-8708},
doi = {https://doi.org/10.1016/j.aim.2019.05.025},
url = {https://www.sciencedirect.com/science/article/pii/S0001870819302671},
author = {Jürgen Jost and Raffaella Mulas}
}

@inproceedings{sinha2015overview,
  title={An overview of microsoft academic service (mas) and applications},
  author={Sinha, Arnab and Shen, Zhihong and Song, Yang and Ma, Hao and Eide, Darrin and Hsu, Bo-June and Wang, Kuansan},
  booktitle={Proceedings of the 24th international conference on world wide web},
  pages={243--246},
  year={2015}
}

@article{wu2021detecting,
  title={Detecting mixing services via mining bitcoin transaction network with hybrid motifs},
  author={Wu, Jiajing and Liu, Jieli and Chen, Weili and Huang, Huawei and Zheng, Zibin and Zhang, Yan},
  journal={IEEE Transactions on Systems, Man, and Cybernetics: Systems},
  volume={52},
  number={4},
  pages={2237--2249},
  year={2021},
  publisher={IEEE}
}

@article{chodrow2020annotated,
  title={Annotated hypergraphs: models and applications},
  author={Chodrow, Philip and Mellor, Andrew},
  journal={Applied network science},
  volume={5},
  number={1},
  pages={9},
  year={2020},
  publisher={Springer}
}

@article{jure2014snap,
  title={SNAP Datasets: Stanford large network dataset collection},
  author={Jure, Leskovec},
  journal={Retrieved December 2021 from http://snap. stanford. edu/data},
  year={2014}
}

@article{landry2024simpliciality,
  title={The simpliciality of higher-order networks},
  author={Landry, Nicholas W and Young, Jean-Gabriel and Eikmeier, Nicole},
  journal={EPJ data science},
  volume={13},
  number={1},
  pages={17},
  year={2024},
  publisher={Springer Berlin Heidelberg}
}

@article{malizia2025hyperedge,
  title={Hyperedge overlap drives explosive transitions in systems with higher-order interactions},
  author={Malizia, Federico and Lamata-Ot{\'\i}n, Santiago and Frasca, Mattia and Latora, Vito and G{\'o}mez-Garde{\~n}es, Jes{\'u}s},
  journal={Nature communications},
  volume={16},
  number={1},
  pages={555},
  year={2025},
  publisher={Nature Publishing Group UK London}
}

@article{lamata2025hyperedge,
  title={Hyperedge overlap drives synchronizability of systems with higher-order interactions},
  author={Lamata-Ot{\'\i}n, Santiago and Malizia, Federico and Latora, Vito and Frasca, Mattia and G{\'o}mez-Garde{\~n}es, Jes{\'u}s},
  journal={Physical Review E},
  volume={111},
  number={3},
  pages={034302},
  year={2025},
  publisher={APS}
}

@article{kraakman2025hypercurveball,
  title={Hypercurveball algorithm for sampling hypergraphs with fixed degrees},
  author={Kraakman, Yanna J and Stegehuis, Clara},
  journal={Journal of Complex Networks},
  volume={13},
  number={4},
  pages={cnaf007},
  year={2025},
  publisher={Oxford University Press}
}

@article{nakajima2021randomizing,
  title={Randomizing hypergraphs preserving degree correlation and local clustering},
  author={Nakajima, Kazuki and Shudo, Kazuyuki and Masuda, Naoki},
  journal={IEEE Transactions on Network Science and Engineering},
  volume={9},
  number={3},
  pages={1139--1153},
  year={2021},
  publisher={IEEE}
}

@article{mancastroppa2023hyper,
  title={Hyper-cores promote localization and efficient seeding in higher-order processes},
  author={Mancastroppa, Marco and Iacopini, Iacopo and Petri, Giovanni and Barrat, Alain},
  journal={Nature communications},
  volume={14},
  number={1},
  pages={6223},
  year={2023},
  publisher={Nature Publishing Group UK London}
}

@inproceedings{genetti2024influence,
  title={Influence maximization in hypergraphs using multi-objective evolutionary algorithms},
  author={Genetti, Stefano and Ribaga, Eros and Cunegatti, Elia and Lotito, Quintino F and Iacca, Giovanni},
  booktitle={International Conference on Parallel Problem Solving from Nature},
  pages={217--235},
  year={2024},
  organization={Springer}
}

@article{saracco2025entropy,
  title={Entropy-based models to randomise real-world hypergraphs},
  author={Saracco, Fabio and Petri, Giovanni and Lambiotte, Renaud and Squartini, Tiziano},
  journal={Communications Physics},
  volume={8},
  number={1},
  pages={284},
  year={2025},
  publisher={Nature Publishing Group UK London}
}

@article{barthelemy2022class,
  title={Class of models for random hypergraphs},
  author={Barthelemy, Marc},
  journal={Physical Review E},
  volume={106},
  number={6},
  pages={064310},
  year={2022},
  publisher={APS}
}

@inproceedings{lee2021hyperedges,
  title={How do hyperedges overlap in real-world hypergraphs?-patterns, measures, and generators},
  author={Lee, Geon and Choe, Minyoung and Shin, Kijung},
  booktitle={Proceedings of the web conference 2021},
  pages={3396--3407},
  year={2021}
}

\begin{widetext}
\newpage
\section*{Supplementary information}
\setcounter{figure}{0}
\setcounter{table}{0}
\setcounter{section}{0}
\renewcommand{\thefigure}{S\arabic{figure}}
\renewcommand{\thetable}{S\arabic{table}}
\renewcommand{\thesection}{S\arabic{section}}

\subsection*{Supplementary Note 1. Datasets}
This section provides detailed descriptions of the datasets used in our experiments. The datasets are originally collected in~\cite{kim2023reciprocity} and represent a diverse range of real-world systems with directed higher-order interactions. Summary statistics of the datasets used in our experiments are reported in Table~\ref{tab:dataset_description}.

\begin{itemize}
    \item \textbf{Question answering data.} We use two \qna datasets: Math-overflow and Server-fault, both sourced from Stack Exchange logs. A hyperedge \( e_{i} = (S_{i}, T_{i}) \) indicates a question posted by the user in the target set \( T_{i} \) and answered by the users in the source set \( S_{i} \). Each hyperedge has a unit target set, i.e., \( |T_{i}| = 1, \forall i = \{1, ..., |E|\} \). If the node in the target set also appears in the source set, we remove it from the source set.
    \item \textbf{Email data.} We use two email datasets: email-enron~\cite{chodrow2020annotated} and email-eu~\cite{jure2014snap}. A hyperedge \( e_{i} = (S_{i}, T_{i}) \) represents an email where the sender is the source set \( S_{i} \), and the receivers (including cc-ed users) form the target set \( T_{i} \). Each hyperedge has a unit source set, i.e., \( |S_{i}| = 1, \forall i = \{1, ..., |E|\} \). If the node in the source set also appears in the target set, we remove it from the target set.
    \item \textbf{Bitcoin transactions data.} We use three bitcoin transaction datasets: bitcoin-2014, bitcoin-2015, and bitcoin-2016~\cite{wu2021detecting}. They contain the first $1\,500\,000$ transactions in 11/2014, 06/2015, and 01/2016 respectively. A hyperedge \( e_{i} = (S_{i}, T_{i}) \) corresponds to a transaction where the accounts from which the coins are sent form the source set \( S_{i} \), and the accounts receiving the coins make up the target set \( T_{i} \).
    \item \textbf{Metabolic data.} We use two methabolic datasets: iAF1260b and iJO1366~\cite{yadati2020nhp}. Nodes are the genes and hyperedges are metabolic reactions. A hyperedge $e_{i} = (S_{i},T_{i})$ indicates that the reaction among genes in the source set $S_{i}$ results in genes in the target set $T_{i}$.
    \item \textbf{Citation data.} We use two citation datasets: citation-data mining and citation-software~\cite{sinha2015overview, yadati2021graph}. A hyperedge \( e_{i} = (S_{i}, T_{i}) \) represents a citation from a paper co-authored by the authors in the source set \( S_{i} \) to a paper co-authored by the authors in the target set \( T_{i} \). Papers with more than 10 authors are filtered out.
\end{itemize}

\begin{table}[h]
    \centering
    \begin{tabular}{lrrrrrrr}
        \toprule
        Dataset & $|V|$ & $|E|$ & $\overline{\left| S_i \right|}$ & $\overline{\left| T_i \right|}$\\
        \midrule
        bitcoin-2014 & 1\,697\,625 & 1\,437\,082 & 1.478 & 1.697\\
        bitcoin-2015 & 1\,961\,886 & 1\,449\,827 & 1.568 & 1.744\\
        bitcoin-2016 & 2\,009\,978 & 1\,451\,135 & 1.495 & 1.715\\
        metabolic-iaf1260b & 1\,668 & 2\,083 & 1.998 & 2.267\\
        metabolic-iJO1366 & 1\,805 & 2\,251 & 2.026 & 2.272\\
        email-enron & 110 & 1\,484 & 1.000 & 2.354\\
        email-eu & 986 & 35\,772 & 1.000 & 2.368\\
        citation-dm & 27\,164 & 73\,113 & 3.253 & 3.038\\
        citation-software & 16\,555 & 53\,177 & 2.927 & 2.717\\
        qna-math & 34\,812 & 93\,731 & 1.779 & 1.000\\
        qna-server & 172\,330 & 272\,116 & 1.747 & 1.000\\
        \bottomrule
    \end{tabular}
    \caption{Summary statistics of the datasets used in our experiments.}
    \label{tab:dataset_description}
\end{table}

\subsection*{Supplementary Note 2. Null model}

\begin{breakablealgorithm}
\caption{Microcanonical set-swap configuration model}
\begin{algorithmic}[1]
\Require Directed hypergraph \(H=(V,E)\) with source sets \(s(e)\) and target sets \(t(e)\); iteration budget \(T\)
\Ensure \(H'\) preserving node in/out degrees and all \(|s(e)|, |t(e)|\)
\State \(H' \gets H\)
\State \(\Sigma \gets \{(s(e),\, t(e)) : e \in E\}\) \Comment{signatures to forbid duplicates}
\For{\(t \gets 1\) \textbf{to} \(T\)}
  \State choose \textit{side} from \{\textit{source}, \textit{target}\}
  \State sample distinct edges \(e_i \neq e_j\) uniformly from \(E\)
  \If{\textit{side} = \textit{source}}
    \State pick \(u \in s(e_i)\), \(v \in s(e_j)\)
    \State \(s'(e_i) \gets (s(e_i)\setminus\{u\}) \cup \{v\}\)
    \State \(s'(e_j) \gets (s(e_j)\setminus\{v\}) \cup \{u\}\)
    \If{\(v \notin s(e_i)\) \textbf{and} \(u \notin s(e_j)\) \textbf{and}
         \(v \notin t(e_i)\) \textbf{and} \(u \notin t(e_j)\) \textbf{and}
         \((s'(e_i), t(e_i)) \notin \Sigma\) \textbf{and} \((s'(e_j), t(e_j)) \notin \Sigma\)}
      \State accept: \(s(e_i)\gets s'(e_i)\); \(s(e_j)\gets s'(e_j)\); update \(\Sigma\)
    \EndIf
  \Else
    \State pick \(u \in t(e_i)\), \(v \in t(e_j)\)
    \State \(t'(e_i) \gets (t(e_i)\setminus\{u\}) \cup \{v\}\)
    \State \(t'(e_j) \gets (t(e_j)\setminus\{v\}) \cup \{u\}\)
    \If{\(v \notin t(e_i)\) \textbf{and} \(u \notin t(e_j)\) \textbf{and}
         \(v \notin s(e_i)\) \textbf{and} \(u \notin s(e_j)\) \textbf{and}
         \((s(e_i), t'(e_i)) \notin \Sigma\) \textbf{and} \((s(e_j), t'(e_j)) \notin \Sigma\)}
      \State accept: \(t(e_i)\gets t'(e_i)\); \(t(e_j)\gets t'(e_j)\); update \(\Sigma\)
    \EndIf
  \EndIf
\EndFor
\State \Return \(H'\);
\end{algorithmic}
\end{breakablealgorithm}

\subsection*{Supplementary Note 3. Additional results about motif analysis in directed hypergraphs}

The combinatorial explosion in the number of possible subhypergraph patterns for a given number of nodes makes it impractical to report statistics for every single motif. In the main text, we therefore only display the visual representations of the most over-represented patterns. Here, we expand this analysis and provide additional quantitative results on the abundance and significance of motifs across datasets.

Figure~\ref{fig:reb_motifs} provides an overview of the structural diversity and frequency distribution of motifs. Panel~(a) reports the number of distinct non-isomorphic subhypergraph patterns with three and four nodes observed across datasets. Although the total number of possible non-isomorphic configurations grows superexponentially with the number of nodes, only a subset of them actually appear in real data. Panel~(b) examines how motif occurrences are distributed among these observed patterns, showing whether they are concentrated in a few motifs or spread more evenly. The results indicate that the top 10\% of motifs account for most occurrences. Some variability is visible across domains; for instance, 3-node motifs in metabolic and email datasets display a slightly more balanced distribution. Overall, motif frequencies are highly concentrated in a relatively small subset of distinct patterns.

\begin{figure*}
    \centering
    \includegraphics[width=\linewidth]{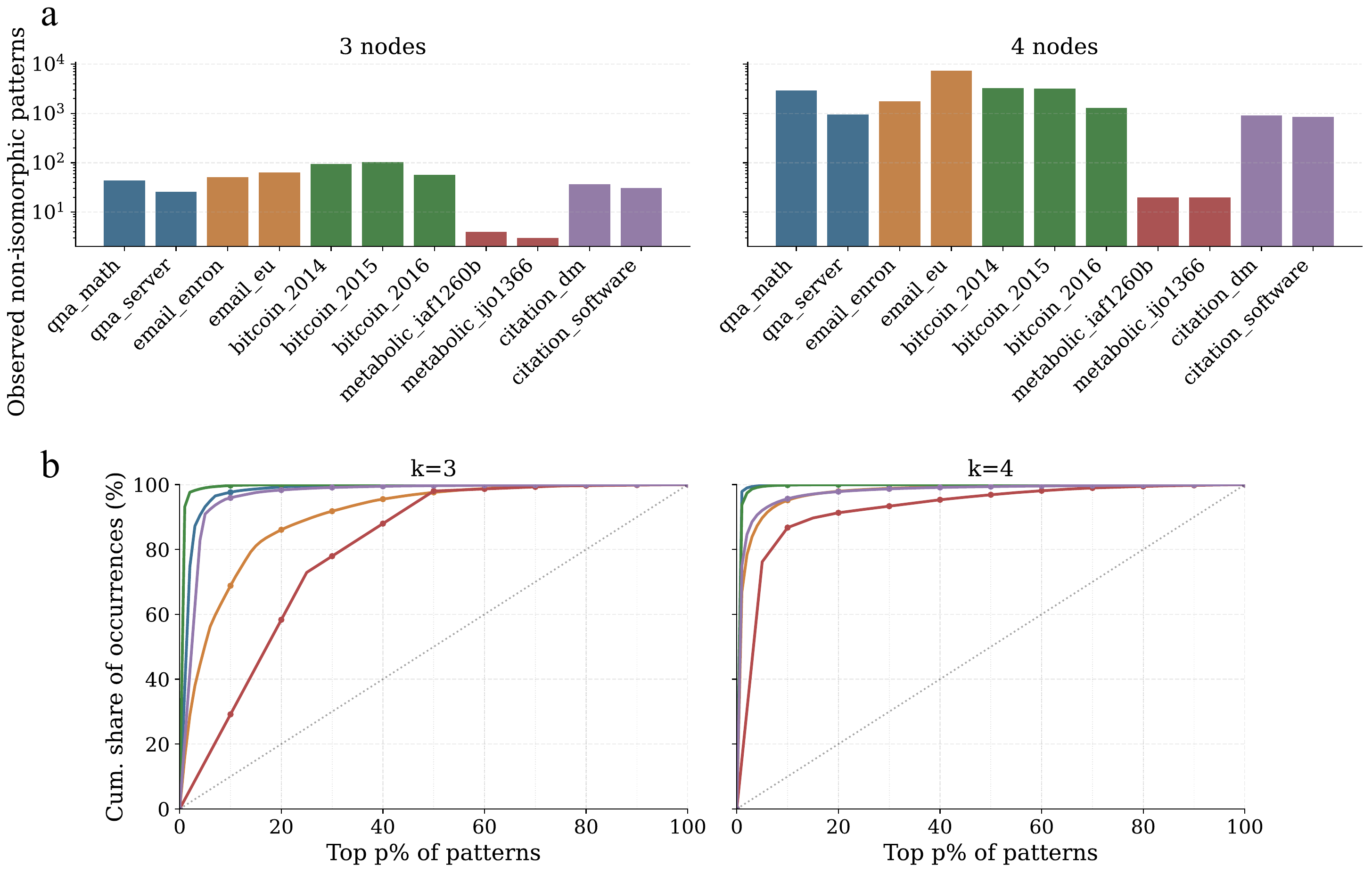}
    \caption{(a) Number of observed non-isomorphic subhypergraph patterns with three and four nodes in each dataset. 
    (b) Cumulative fraction of motif occurrences versus the ranked motif index, showing that few motifs account for most occurrences.}
    \label{fig:reb_motifs}
\end{figure*}

To assess the statistical significance of these motifs, we compute the z-scores (see Methods) and analyze their distribution, shown in Fig.~\ref{fig:reb_motifs2}. The distributions, capped between $-10$ and $10$, reveal the range of significance levels across datasets. The motif drawings presented in the main paper (see Fig. $7$) correspond to those lying in the positive tail of the distributions, that is, motifs whose observed frequencies deviate by several standard deviations from the null expectation.

\begin{figure*}
    \centering
    \includegraphics[width=\linewidth]{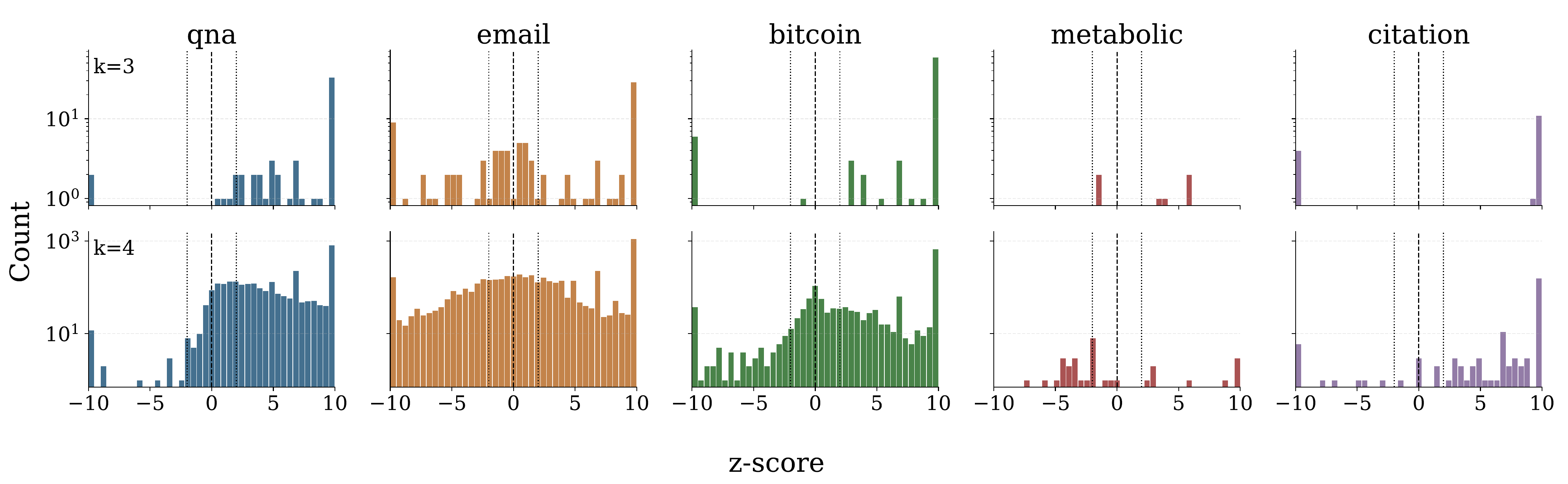}
    \caption{Distribution of motif z-scores across datasets, capped between $-10$ and $10$. 
    Motifs shown in the main paper correspond to those in the positive tail of the distribution, indicating strong over-representation relative to the configuration model.}
    \label{fig:reb_motifs2}
\end{figure*}

\subsection*{Supplementary Note 4. Algorithms for motif analysis in directed hypergraphs}
This section provides further details on the algorithms for motif analysis in directed hypergraphs. In Table~\ref{tab:exec_times}, we report the execution times of our algorithms for motifs of order $3$ and $4$ across various datasets, highlighting the increase in time when moving from order $3$ to order $4$. More complex approaches will be needed to scale the analysis to larger motifs and larger dataset sizes. In Algorithm~\ref{alg:motifs_order3} and Algorithm~\ref{alg:motifs_order4}, we present detailed pseudocode for the algorithms designed to count directed higher-order motifs of sizes $3$ and $4$, respectively.

\begin{table}
\caption{Execution times by dataset for Order 3 and Order 4 (mean and standard deviation, in seconds).}
\label{tab:exec_times}
\centering
\sisetup{
  table-number-alignment = center,
  group-separator = {\,},
  group-minimum-digits = 4
}
\setlength{\tabcolsep}{4pt}
\begin{tabular}{
l
S[table-format=4.2] @{\,\(\pm\)\,} S[table-format=2.2]
S[table-format=4.2] @{\,\(\pm\)\,} S[table-format=2.2]
}
\toprule
& \multicolumn{2}{c}{Order 3 (s)} & \multicolumn{2}{c}{Order 4 (s)} \\
\cmidrule(lr){2-3} \cmidrule(lr){4-5}
Dataset & {Mean} & {Std} & {Mean} & {Std} \\
\midrule
bitcoin-2014         & 26.75 & 1.75 & 2217.30 & 41.26 \\
bitcoin-2015         & 24.58 & 1.09 & 1507.48 & 5.80  \\
bitcoin-2016         & 28.05 & 1.39 & 5109.27 & 35.94 \\
citation-dm          & 0.43  & 0.01 & 19.79   & 0.17  \\
citation-software    & 0.42  & 0.00 & 22.06   & 0.07  \\
email-enron          & 0.03  & 0.00 & 3.27    & 0.03  \\
email-eu             & 0.73  & 0.00 & 400.91  & 0.63  \\
metabolic-iaf1260b   & 0.01  & 0.00 & 0.24    & 0.00  \\
metabolic-ijo1366    & 0.01  & 0.00 & 0.27    & 0.01  \\
qna-math             & 1.39  & 0.02 & 1210.31 & 2.84  \\
qna-server           & 3.86  & 0.17 & 4918.33 & 20.80 \\
\bottomrule
\end{tabular}
\end{table}

\clearpage
\begin{breakablealgorithm}
\caption{Motifs of order 3}\label{alg:motifs_order3}
\textbf{Input:} A directed hypergraph $\mathcal{H}=(V,E)$

\textbf{Output:} distribution of the frequency of the motifs of order 3
\begin{algorithmic}[1]
\State Let M be the motifs frequency hash map
\State Let U be the isomorphism class hash map
    \ForEach{hyperedge $e$ of order 3 in $E$}
        \State $V^{*} \gets$ vertices of $e$
        \State $\mathit{motif} \gets \emptyset $
        \ForEach{ $ e^{*} \in \mathcal{P}(V^{*}) $ }
            \If{ $e^{*} \in E$ }
                \State $\mathit{motif} \gets \mathit{motif} \cup e^{*} $
            \EndIf
        \EndForEach
        \State $C_m \gets$ lexicographically minimum canonic relabel of $\mathit{motif}$

        \If{$C_m \notin M$}
            \State $M[C_m] \gets 0$
        \EndIf
        \State $M[C_m] += 1$ 
        \State Set vertices of $\mathit{motif}$ as visited
          
    \EndForEach
\end{algorithmic}
\end{breakablealgorithm}

\newpage
\begin{breakablealgorithm}
\caption{Motifs of order 4}\label{alg:motifs_order4}
\textbf{Input:} A directed hypergraph $H=(V,E)$

\textbf{Output:} distribution of the frequency of the motifs of order 4
\begin{algorithmic}[1]
\State Let $M$ be the motifs frequency hash map
\State Let $s$ be the isomorphism class hash map
    \ForEach{hyperedge $e$ of order 4 in $E$}
        \State $V^{*} \gets$ vertices of $e$
        \State $\mathit{motif} \gets \emptyset $
        \ForEach{ $ e^{*} \in \mathcal{P}(V^{*}) $ }
            \If{ $e^{*} \in E$ }
                \State $\mathit{motif} \gets \mathit{motif} \cup e^{*} $
            \EndIf
        \EndForEach
        \State $C_m \gets$ lexicographically minimum canonic relabel of $\mathit{motif}$

        \If{$C_m \notin M$}
            \State $M[C_m] \gets 0$
        \EndIf
        
        \State $M[C_m] += 1$  
        \State Set vertices of $\mathit{motif}$ as visited
    \EndForEach
    \State $\mathcal{H} \gets $Discard all hyperedges of order 4 from $\mathcal{H}$
    \ForEach{hyperedge $e$ of order 3 in $E$}
        \State Let $\mathcal{Z}$ be the set of hyperedges adjacent to $e$
        \ForEach{hyperedge $ \zeta $ in $ \mathcal{Z}$ }
        \If{ $|\zeta \cup e|=4$ and $\zeta \cup e$ not already visited }
            \State $V^{*} \gets$ vertices of $\zeta \cup e$
            \State $\mathit{motif} \gets \emptyset $
            \ForEach{ $ e^{*} \in \mathcal{P}(V^{*}) $ }
                \If{ $e^{*} \in E$ }
                    \State $\mathit{motif} \gets \mathit{motif} \cup e^{*} $
                \EndIf
            \EndForEach
            \State $C_m \gets$ lexicographically minimum canonic relabel of $\mathit{motif}$
    
            \If{$C_m \notin M$}
                \State $M[C_m] \gets 0$
            \EndIf
            
            \State $M[C_m] += 1$ 
            \State Set vertices of $ \mathit{motif} $ as visited
        \EndIf
        \EndForEach
    \EndForEach
\end{algorithmic}
\end{breakablealgorithm}

\newpage
\subsection*{Supplementary Note 5. Graphical outlines of the algorithms proposed}
Here, we provide a schematic flowchart representation of all algorithmic components introduced in the main text.
The diagrams summarize the logical structure, decision steps, and computational complexity of each procedure.
They are intended to complement the formal descriptions and pseudocode, highlighting the conceptual flow of the algorithms
used for the set-swap configuration model (Fig.~\ref{fig:flow_chart_null}), reciprocity evaluation (Fig.~\ref{fig:flow_chart_rec}) and motif discovery (Fig.~\ref{fig:flow_chart_motifs}).

\begin{figure*}[b]
    \centering
\includegraphics[width=0.4\linewidth]{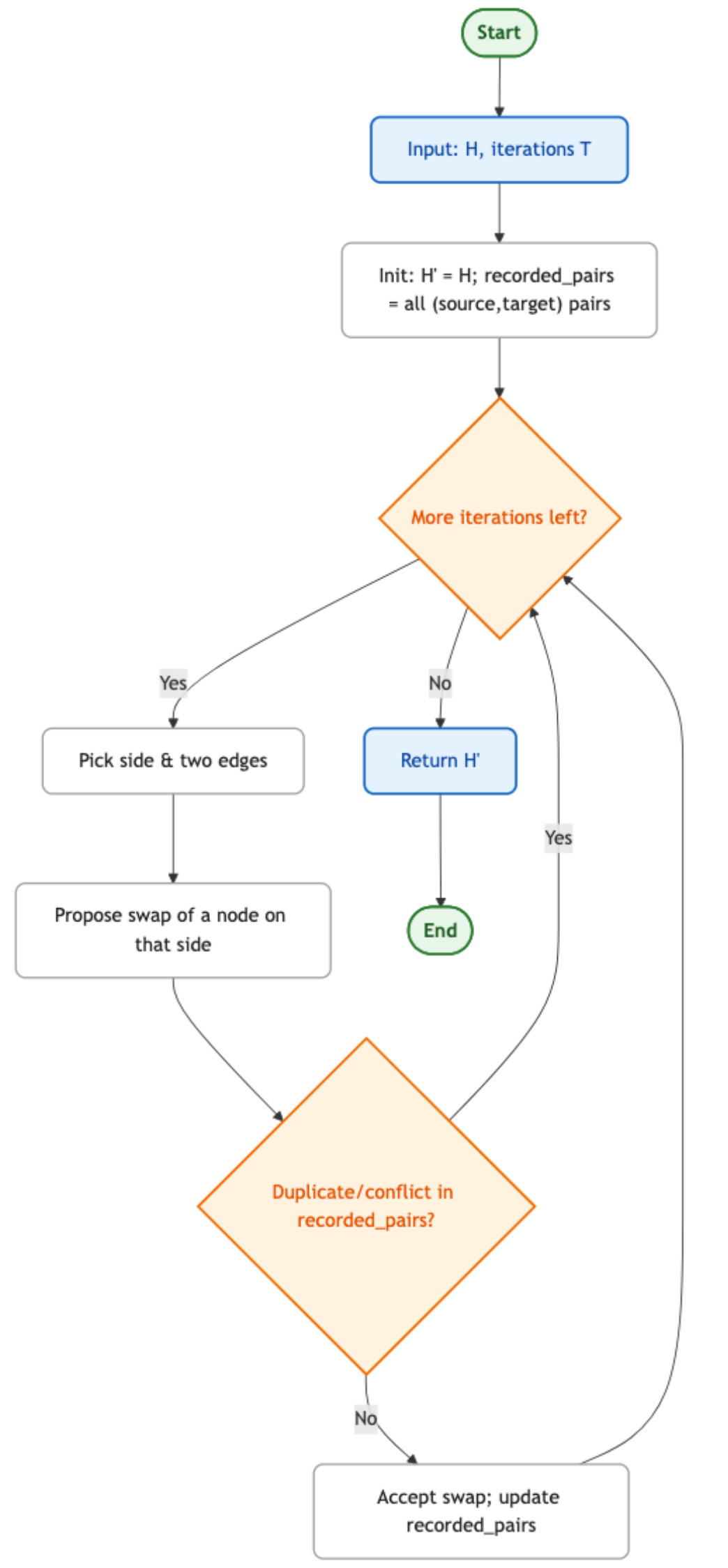}
\caption{\textbf{Microcanonical set-swap configuration model.}
Starting from an observed hypergraph $H=(V,E)$, two edges are randomly selected at each iteration, and one vertex is swapped
between their source or target sets. The swap is accepted only if it does not produce duplicates.
Repeating this process $T$ times preserves node in/out-degrees and edge sizes, generating randomized but structurally
consistent hypergraphs.}
    \label{fig:flow_chart_null}
\end{figure*}

\begin{figure*}
    \centering
    \includegraphics[width=\linewidth]{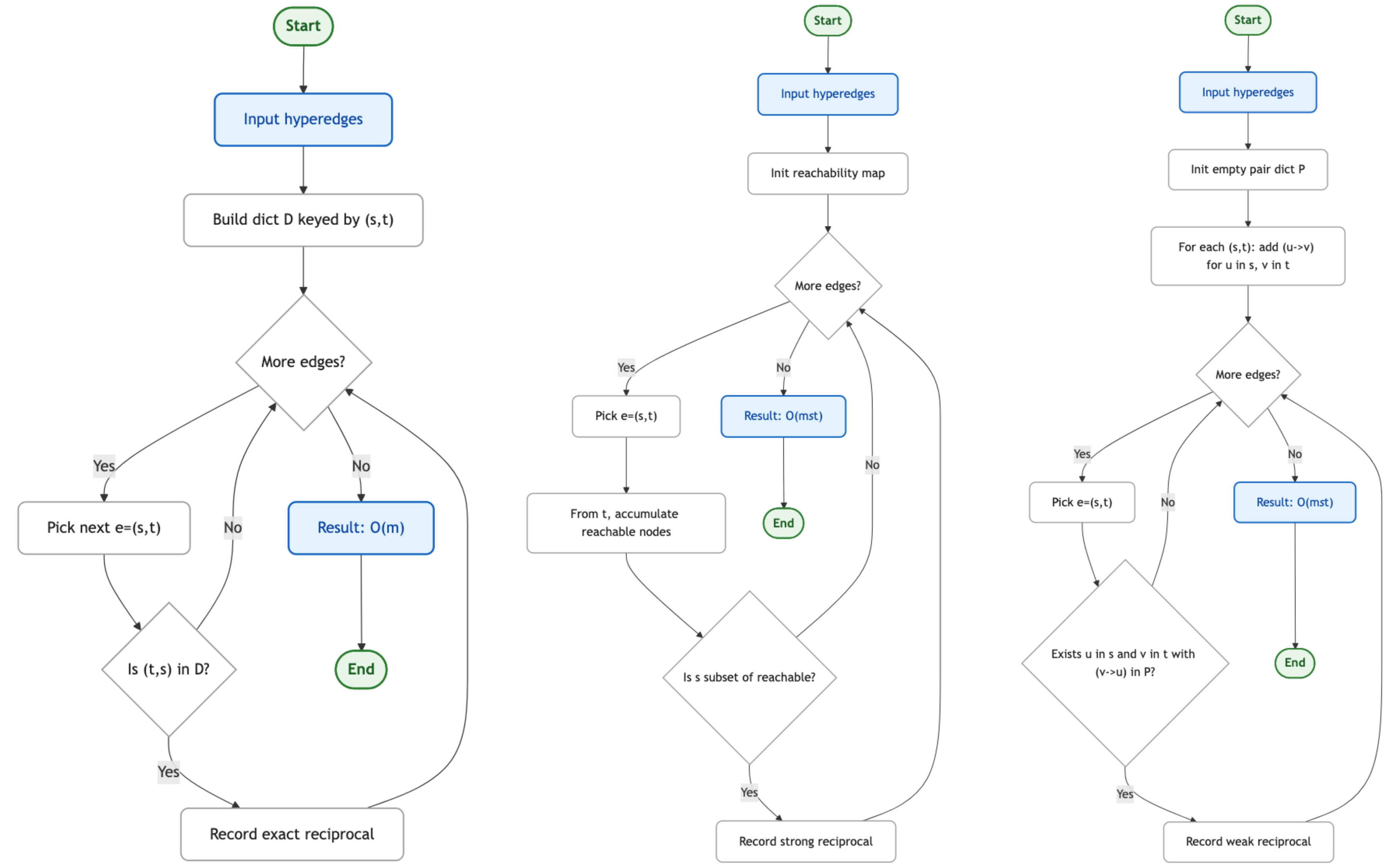}
    \caption{\textbf{Reciprocity algorithms.}
On the left, the exact reciprocity algorithm stores each hyperedge and checks for its reverse.
In the middle, the strong reciprocity algorithm accumulates nodes reachable from the target set to verify full coverage of the source set.
On the right, the weak reciprocity algorithm builds all source–target pairs and checks for any reverse target–source connection.}

    \label{fig:flow_chart_rec}
\end{figure*}

\begin{figure*}
    \centering
    \includegraphics[width=\linewidth]{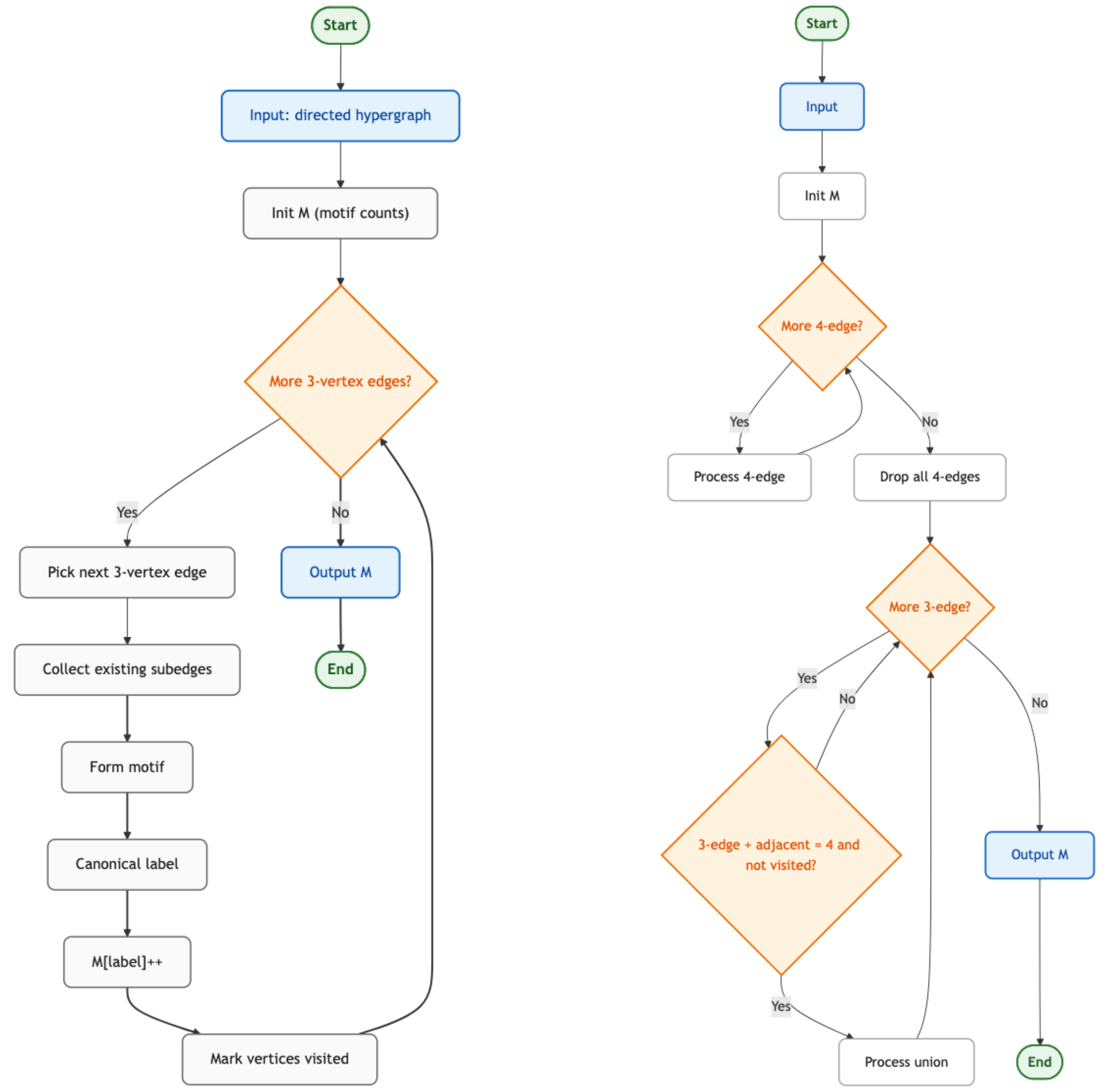}
    \caption{\textbf{Motif discovery algorithms.}
On the left, order-3 motifs are identified from each 3-vertex hyperedge by collecting and relabelling connected subhyperedges.
In the middle, order-4 motifs are extracted from single 4-vertex hyperedges using the same procedure.
On the right, additional order-4 motifs arise from pairs of adjacent 3-vertex hyperedges spanning four vertices.}

    \label{fig:flow_chart_motifs}
\end{figure*}

\end{widetext}

\end{document}